\documentclass[11pt,a4paper]{article}

\usepackage[margin=1in]{geometry}
\usepackage[T1]{fontenc}
\usepackage[utf8]{inputenc}
\usepackage{lmodern}
\usepackage{microtype}

\usepackage{amsmath}
\usepackage{amssymb}
\usepackage{amsfonts}
\usepackage{amsthm}
\usepackage{bm}
\usepackage{dsfont}

\usepackage{graphicx}
\usepackage{epstopdf}
\usepackage{subfigure}
\usepackage{float}
\usepackage{xcolor}
\usepackage{booktabs}
\usepackage{array}
\usepackage{multirow}
\usepackage{ragged2e}
\usepackage[flushleft]{threeparttable}

\usepackage{tikz}
\usetikzlibrary{
    positioning,
    arrows.meta,
    shapes.geometric
}

\usepackage[round,authoryear]{natbib}
\usepackage{url}
\usepackage[
    colorlinks=true,
    citecolor=blue,
    linkcolor=blue,
    urlcolor=blue
]{hyperref}

\theoremstyle{plain}

\theoremstyle{definition}

\theoremstyle{remark}

\begin{document}

\title{
Measuring intraday return-shape geometry
with modified Brownian motion
}

\author{
Raffaele Mattera\textsuperscript{1,*},
Rajabali Ghasem Pour\textsuperscript{1},
Parmjit Kaur\textsuperscript{2},\\
Valerio Ficcadenti\textsuperscript{3},
Gurjeet Dhesi\textsuperscript{4,5}
\\[1em]
\small \textsuperscript{1} University of Campania ``Luigi Vanvitelli'', Italy\\
\small \textsuperscript{2} Babes-Bolyai University, Romania\\
\small \textsuperscript{3} London South Bank University, United Kingdom\\
\small \textsuperscript{4} Nanjing University of Aeronautics and Astronautics, China\\
\small \textsuperscript{5} Bucharest University of Economic Studies, Romania\\[0.5em]
\small \textsuperscript{*} Corresponding author: 
\href{mailto:raffaele.mattera@unicampania.it}{raffaele.mattera@unicampania.it}
}
\date{}

\maketitle

\begin{abstract}
\noindent Realised kurtosis measures intraday tail risk but compresses the shape of the return distribution into a single scalar, so trading days with similar kurtosis may differ in how probability mass is spread across the peak, the shoulders and the tails. We propose a geometric description of intraday return shape based on the modified Brownian motion model (MBMM), which deforms a Gaussian benchmark through a nonlinear feedback governed by a shape parameter and a distortion intensity. Our main contribution is a peak-shoulder-tail decomposition that measures where the fitted return density gains or loses probability mass relative to the Gaussian, complemented by two feedback landmarks that explain how the shape parameter generates these deformations. We estimate the model at the stock-day level using a bin-free simulation-based minimum-distance estimator that matches empirical and model-implied intraday quantiles, and we document its finite-sample reliability by Monte Carlo. Using regular session one-minute returns for ten liquid U.S. stocks from May 2021 to April 2026, we find that the MBMM provides a substantially better in-sample fit than geometric Brownian motion on every estimator-ready stock-day, and that days with closely matched realised kurtosis can display materially different peak-shoulder-tail geometries. Finally, we show that the estimated shape parameters behave as persistent daily state variables: they are strongly forecastable relative to a random-walk benchmark, though their predictability reflects persistence around a slowly moving level rather than rich autoregressive dynamics. The proposed decomposition offers a risk-shape measurement layer that complements realised kurtosis by describing not only how large intraday tail risk is, but how it is formed, giving investors and risk managers additional distributional information for risk monitoring and, potentially, prediction.
\end{abstract}



\section{Introduction}
\label{sec:introduction}

The kurtosis of financial returns has long been recognised as an economically relevant feature of the return distribution. It affects the probability assigned to large gains and losses, enters portfolio and risk-management decisions, and provides information that is not contained in the conditional mean and variance alone \citep{harvey2010portfolio,jondeau2006optimal}. A broad literature therefore studies higher moments as time-varying objects, develops conditional models for skewness and kurtosis \citep{jondeau2003conditional,leon2005autoregresive}, and investigates their implications for asset pricing, volatility, risk measurement, and portfolio allocation \citep{braga2023kurtosis,liu2020effectiveness}. The increasing availability of high-frequency data has extended this perspective from latent conditional moments to \textcolor{black}{co-kurtosis \citep[e.g.][]{bae2021realized,dai2021multiscale} and} realised daily measures constructed from intraday returns \citep{AmayaChristoffersenJacobsVasquez2015,ahadzie2020effects}. In this literature, realised kurtosis is typically analysed as a daily state variable whose dynamics, persistence, and forecastability may be modelled in the same spirit as realised volatility.

This paper asks a different question. Existing work studies extensively how kurtosis changes over time \citep{jondeau2012importance}, how it can be predicted, and whether it improves forecasts of future risk. Much less attention has been paid to the distributional geometry represented by a given realised-kurtosis value. Kurtosis is informative, but necessarily compressive: it maps the shape of an entire intraday return distribution into a single scalar. Consequently, two trading days may display very similar realised kurtosis while differing in the way probability mass is distributed across the \textcolor{black}{peak}, the shoulders, and the tails. 

\begin{figure}[!htbp]
\centering
\includegraphics[width=\textwidth]{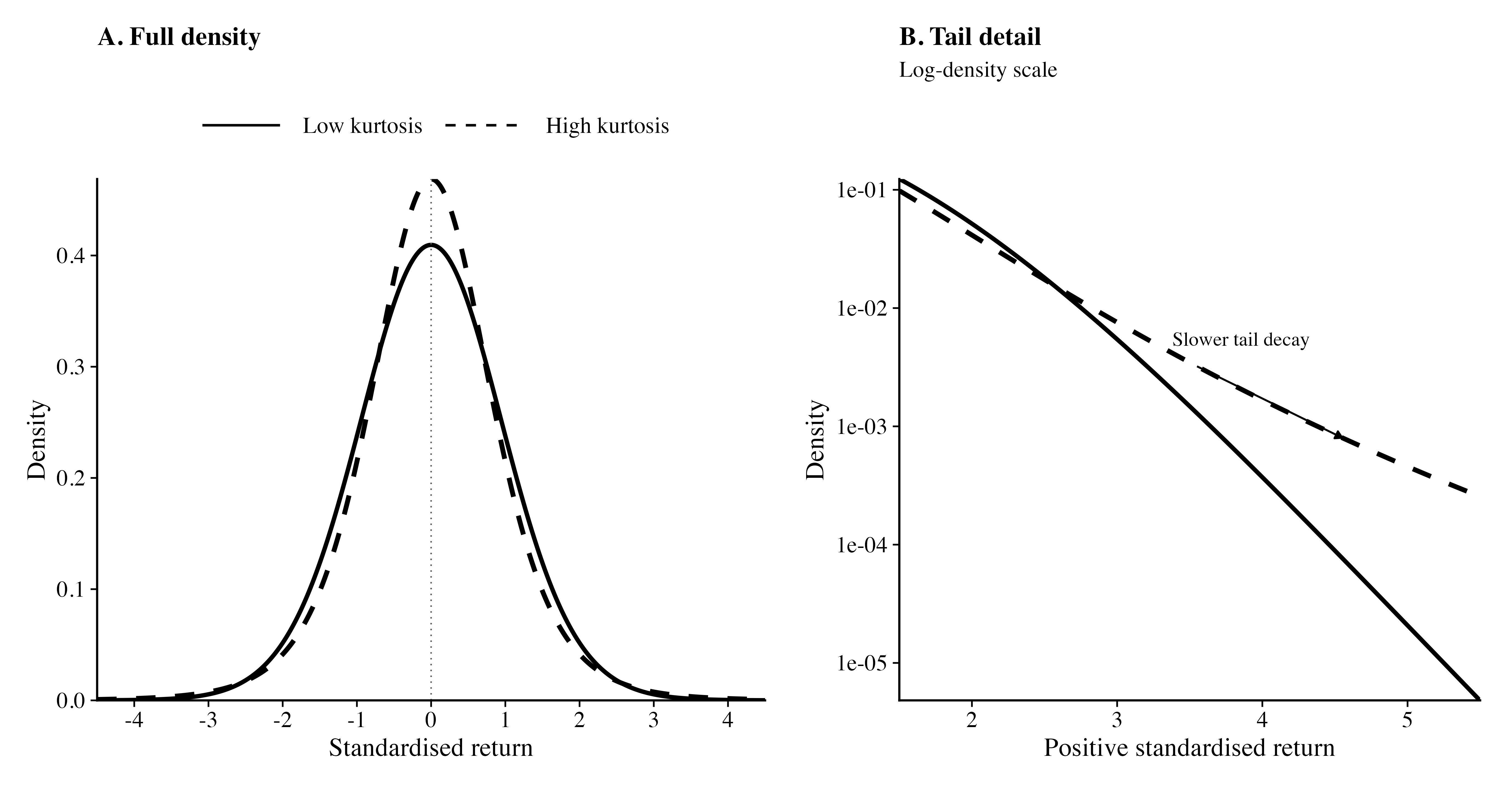}
\caption{Comparison of two symmetric standardised return distributions with common mean
zero, unit variance, and zero skewness, but different kurtosis. The
low-kurtosis curve has kurtosis \(3.23\), whereas the high-kurtosis curve has
kurtosis \(6\). The higher-kurtosis distribution displays stronger central
concentration and heavier tails, illustrating the type of shape variation
summarised by the fourth moment.}
\label{fig:kurtosis_only_comparison}
\end{figure}

Figure~\ref{fig:kurtosis_only_comparison} illustrates the starting point of our argument. The two displayed distributions are symmetric, have mean zero and unit variance, and differ only in kurtosis.  
The figure shows the familiar distributional effect summarised by the fourth standardised moment. However, one can believe that two distributions with similar $\kappa$ may have the same shape. Our contribution therefore begins from the next question: even after kurtosis has been measured, can distributions with very similar realised kurtosis conceal materially different internal geometries? 

To investigate this question in a financially meaningful setting, we work with realised kurtosis rather than with an abstract population moment. The use of one-minute returns allows us to estimate this quantity separately for each stock-day and, crucially, to associate each daily realised-kurtosis observation with a daily estimate of the underlying distributional geometry. The resulting object is not therefore a generic population-level description of kurtosis, but a daily realised geometry of intraday returns.

A natural starting point for constructing this geometry is the geometric Brownian motion model. Despite its limitations, geometric Brownian motion remains the canonical benchmark for financial-price dynamics and provides a transparent Gaussian reference distribution. To account for kurtosis, we consider its extension \textcolor{black}{appeared in the initial investigations of \citet{DhesiShakeelXiao2016,dhesi2016modelling} and \citet{DhesiShakeelAusloos2021}, the modified Brownian motion model (MBMM)}. Non-Gaussian extensions of Brownian price dynamics have also been developed to reproduce empirically observed return shapes and tail behaviour. For example, \cite{shaw2015model} introduce feedback into Brownian dynamics and obtain a rich class of fat-tailed return distributions. MBMM takes a different route, deforming the Gaussian innovation directly through the nonlinear feedback function \(f_c\), which itself depends on two key parameters. The parameter \(c\) governs the shape of the feedback mechanism, while the coefficient \(K\) governs the strength and direction of the distortion. This structure makes MBMM particularly suitable for our purpose. Indeed, instead of merely reporting that a return distribution is non-Gaussian or leptokurtic, MBMM provides a parametric description of how the Gaussian benchmark is deformed.

Our primary contribution is a peak--shoulder--tail decomposition of the intraday return distribution. Rather than summarising the deformation of the Gaussian benchmark through a single scalar, the decomposition measures directly \emph{where} the fitted MBMM density gains or loses probability mass: a central-excess region, an intermediate-depletion region, and an outer-tail-excess region. This yields a model-implied geometric representation of how probability mass is redistributed relative to the Gaussian benchmark. It is useful as a financial diagnostic because it distinguishes the magnitude of realised tail risk from the mechanism through which that risk is formed. To interpret this mechanism, we additionally characterise two landmarks induced by the shape parameter \(c\) on the latent innovation scale. The first, \(z_P(c)\), identifies the location at which the nonlinear feedback reaches its strongest interior effect; the second, \(z_T(c)\), identifies the point at which the feedback changes sign and the transformation enters its outer transition region. These landmarks do not themselves define the peak, shoulder, and tail regions, which are obtained from the fitted return density; rather, they describe how the feedback deforms the Gaussian innovation and thereby help explain why different values of \(c\) generate different peak--shoulder--tail geometries.


A second contribution concerns the estimation of these parameters of interest. Existing MBMM applications calibrate the model through simulation-based goodness-of-fit procedures applied to binned return distributions \citep{DhesiShakeelXiao2016,DhesiShakeelAusloos2021}. Although intuitive, histogram-based fitting depends on bin construction and is not naturally suited to repeated daily estimation from intraday data. We therefore adopt a simulation-based minimum-distance (SMDE) estimator that matches empirical and model-implied intraday quantiles. For each candidate pair \((c,\lambda)\), the estimator simulates MBMM return samples, computes their quantiles, and minimises a quadratic distance from the empirical quantile vector. The procedure is bin-free, directly targets the distributional features of interest, and makes daily MBMM estimation operational.

The third contribution is of empirical nature. Using regular-session one-minute returns for ten liquid U.S. stocks from May 2021 to April 2026, we estimate the MBMM parameters and realised kurtosis for each estimator-ready stock-day. We document that realised kurtosis and MBMM geometry are strongly related but not equivalent: a non-negligible subset of stock-days with closely matched realised kurtosis displays materially different peak--shoulder--tail allocations and different feedback landmarks. Hence, we show how realised kurtosis captures the overall magnitude of fourth-moment risk, whereas MBMM supplies a richer description of its fitted internal shape.

This distinction is relevant for investors and risk managers. Two days with comparable realised kurtosis need not imply the same operational risk environment. A geometry in which the transition region lies close to the centre may require different monitoring, exposure limits, or hedging attention from a geometry in which tail-like behaviour begins only at more extreme standardised innovations. We do not interpret the MBMM geometry as a direct trading signal or as evidence of profitability. Rather, it provides a risk-shape measurement layer that may complement realised volatility, realised kurtosis, Value-at-Risk, Expected Shortfall, and option-implied indicators.


The remainder of the paper is organised as follows. Section~\ref{sec:mbmm_geometry} introduces MBMM, defines the peak--shoulder--tail decomposition of the return distribution, and characterises the feedback landmarks induced by \(c\). Section~\ref{sec:estimation} presents the simulation-based minimum-distance estimator, while Section~\ref{sec:simulation} studies its finite-sample behaviour. Section~\ref{sec:data} describes the one-minute return data and the construction of daily realised kurtosis. Section~\ref{sec:insample_fit} presents the in-sample geometric evidence, and Section~\ref{sec:forecasting_tail_indicators} examines the temporal predictability of the MBMM-derived indicators. Section~\ref{sec:conclusions} concludes the paper with final remarks and suggestions for future research.

\section{Modified Brownian motion and return-shape geometry}
\label{sec:mbmm_geometry}

\subsection{From geometric to modified Brownian motion}
\label{subsec:gbm_to_mbmm}

Let \(P_t\) denote the asset price at time \(t\), and let
\(Z_t\overset{\mathrm{iid}}{\sim}\mathcal N(0,1)\). Over an interval
of length \(\Delta\), the geometric Brownian motion benchmark is
written as
\begin{equation}
P_{t+\Delta}
=
P_t
\exp\left(
\mu\Delta+\sigma\sqrt{\Delta}\,Z_t
\right),
\label{eq:gbm_price}
\end{equation}
where \(\mu\) is the drift and \(\sigma>0\) is the volatility scale.
The corresponding log return is conditionally Gaussian. Although this
benchmark remains foundational in financial modelling, its Gaussian
innovation cannot reproduce the pronounced central concentration and
tail behaviour commonly observed in intraday financial returns.

Following \citet{DhesiShakeelXiao2016} and
\citet{DhesiShakeelAusloos2021}, the modified Brownian motion model
augments the Gaussian component with a nonlinear feedback term:
\begin{equation}
P_{t+\Delta}
=
P_t
\exp\left[
\mu\Delta
+
\sigma\sqrt{\Delta}\,Z_t
+
\lambda f_c(Z_t)\Delta
\right],
\qquad
\lambda=\mu K,
\label{eq:mbmm_price}
\end{equation}
where
\begin{equation}
f_c(z)
=
\left[
2\exp\left(-\frac{cz^2}{2}\right)-1
\right]
\arctan(z),
\qquad
c>0.
\label{eq:feedback_function}
\end{equation}
The reparametrisation \(\lambda=\mu K\) is particularly useful for
intraday estimation because it avoids attempting to recover
\(K=\lambda/\mu\) when the daily intraday mean return is close to zero. The corresponding return representation is
\begin{equation}
r_t
=
\mu\Delta
+
\sigma\sqrt{\Delta}\,Z_t
+
\lambda f_c(Z_t)\Delta.
\label{eq:mbmm_return}
\end{equation}
For notational simplicity, we henceforth set \(\Delta=1\). After centring and scaling by the Gaussian volatility parameter,
\begin{equation}
X_t
=
\frac{r_t-\mu}{\sigma}
=
Z_t+\alpha f_c(Z_t),
\qquad
\alpha=\frac{\lambda}{\sigma}.
\label{eq:standardised_mbmm}
\end{equation}
Thus, the standardised MBMM return is generated by applying the
nonlinear transformation
\begin{equation}
h_{c,\alpha}(z)
=
z+\alpha f_c(z)
\label{eq:mbmm_transformation}
\end{equation}
to a standard-normal innovation. The feedback function \(f_c\) is not itself a probability density.
Instead, it determines how each Gaussian innovation \(z\) is displaced
before the return distribution is formed. The shape of the resulting
distribution depends jointly on the geometry of \(f_c\), the shape
parameter \(c\), and the magnitude and sign of \(\alpha\).

\begin{figure}[!htbp]
\centering
\includegraphics[width=0.94\textwidth]
{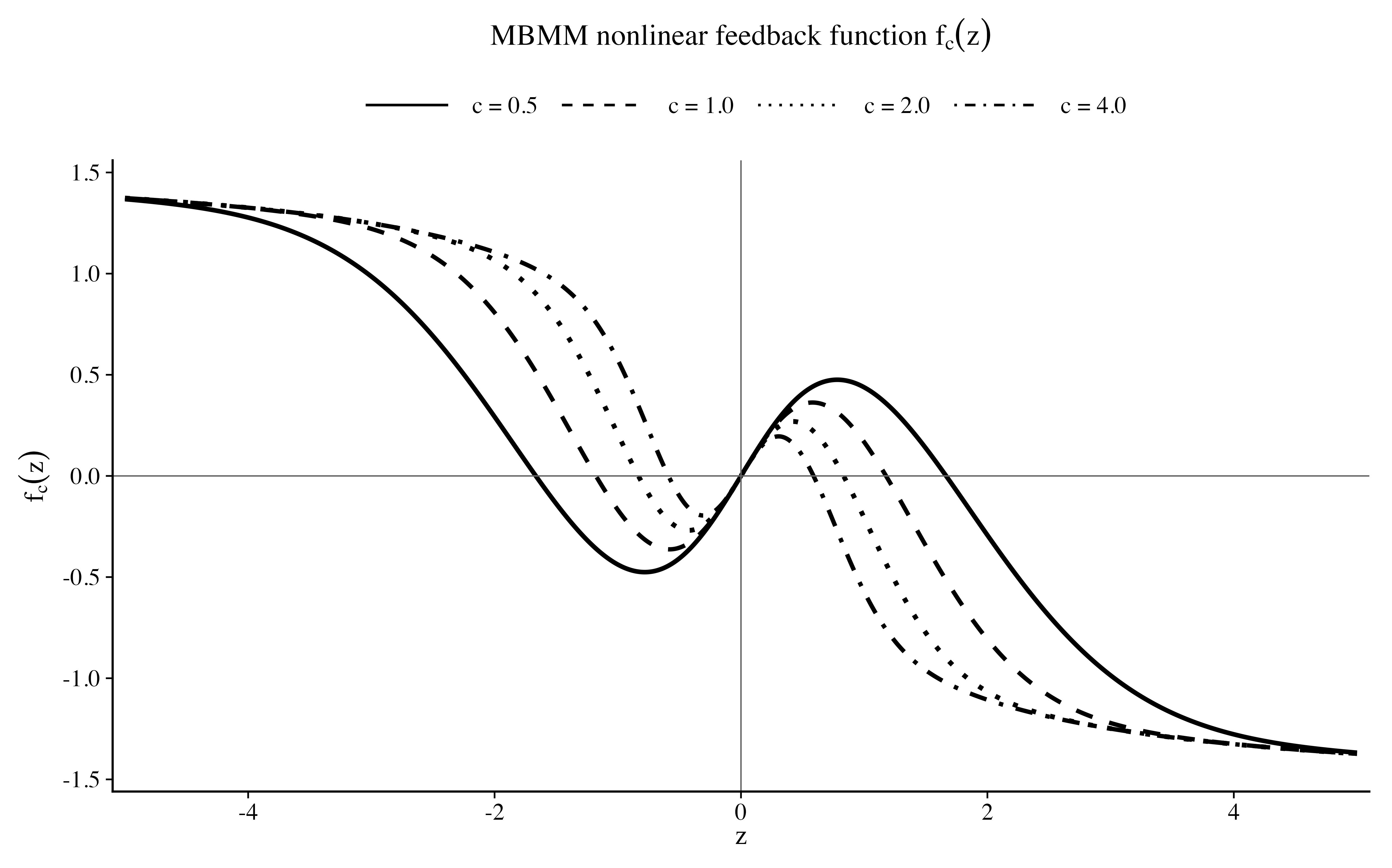}
\caption{
Nonlinear MBMM feedback function \(f_c(z)\) for selected values of
\(c\). The parameter \(c\) controls the horizontal geometry of the
feedback function and, in particular, the locations of its extrema
and non-zero roots. Smaller values of \(c\) move these features farther
from the origin, whereas larger values move them towards the centre.
}
\label{fig:mbmm_feedback_function}
\end{figure}

Figure~\ref{fig:mbmm_feedback_function} illustrates the feedback
function for selected values of \(c\). 
Since \(f_c\) is odd, positive
and negative innovations receive corrections of opposite signs.
Changing \(c\) modifies the horizontal geometry of the function,
including the positions of its internal extrema and non-zero roots.
In contrast, \(\alpha\) controls the magnitude and direction of the
actual correction \(\alpha f_c(z)\). The parameters \(c\) and \(\alpha\) therefore have distinct but
interacting roles. The parameter \(c\) controls where the nonlinear
feedback reaches its internal extrema and where it changes sign. The
absolute value \(|\alpha|\) controls the strength of the displacement,
whereas the sign of \(\alpha\) determines its direction. 

Neither
parameter is itself a measure of kurtosis. Rather, kurtosis is a global
distributional implication of their joint action. Let
\begin{equation*}
X
=
Z+\alpha f_c(Z),
\qquad
Z\sim\mathcal N(0,1).
\end{equation*}
Since \(f_c\) is odd and the distribution of \(Z\) is symmetric, we have $\mathbb E[X]=0$. The second moment of the transformed innovation is
\begin{align}
m_2(c,\alpha)
&=
\mathbb E[X^2]=
\nonumber\\
&=
1
+
2\alpha
\mathbb E\left[
Zf_c(Z)
\right]
+
\alpha^2
\mathbb E\left[
f_c(Z)^2
\right].
\label{eq:mbmm_second_moment}
\end{align}
The corresponding fourth moment is
\begin{align}
m_4(c,\alpha)
&=
\mathbb E[X^4]
\nonumber\\
&=
3
+
4\alpha
\mathbb E\left[
Z^3f_c(Z)
\right]
+
6\alpha^2
\mathbb E\left[
Z^2f_c(Z)^2
\right]
\nonumber\\
&\quad
+
4\alpha^3
\mathbb E\left[
Zf_c(Z)^3
\right]
+
\alpha^4
\mathbb E\left[
f_c(Z)^4
\right].
\label{eq:mbmm_fourth_moment}
\end{align}
The model-implied kurtosis is therefore
\begin{equation}
\kappa(c,\alpha)
=
\frac{
m_4(c,\alpha)
}{
m_2(c,\alpha)^2
}.
\label{eq:mbmm_kurtosis}
\end{equation}
Equation~\eqref{eq:mbmm_kurtosis} shows that kurtosis depends jointly
and nonlinearly on \(c\) and \(\alpha\). The parameter \(c\) determines
where the transformation compresses or amplifies innovations, whereas
\(\alpha\) determines the intensity and direction of these
displacements. Different parameter pairs may consequently imply
similar kurtosis values while producing different peak, shoulder, and
tail geometries. When \(\lambda=0\), and therefore \(\alpha=0\), the nonlinear
correction vanishes, so the model then reduces to the Gaussian GBM benchmark, $X=Z,$ and its kurtosis is $\kappa(c,0)
=
\mathbb E[Z^4]
=
3.$ In this case, \(c\) has no effect on the return distribution and is
not identified.

More generally, the MBMM distribution is leptokurtic relative to the
Gaussian benchmark when 
$m_4(c,\alpha)
>
3m_2(c,\alpha)^2$, platykurtic when $
m_4(c,\alpha)
<
3m_2(c,\alpha)^2$
and mesokurtic when equality holds. The sign of \(\alpha\) alone is not sufficient to determine whether
\(\kappa(c,\alpha)\) is above or below Gaussian kurtosis, because the nonlinear
term changes both the fourth moment and the variance used to
standardise it. The direction in which kurtosis departs from "three"
depends jointly on the sign of \(\alpha\) and on \(c\). The relationship between the MBMM parameters and kurtosis is therefore
structural, but kurtosis remains only a scalar summary of the full
distributional deformation. \cite{chen2022dynamic}, for example, represent daily intraday-return distributions through quantile functions and model their dynamics directly. Our perspective is complementary but different. Rather than forecasting the full quantile function, we ask whether a parsimonious stochastic deformation of the Gaussian benchmark can reveal where probability mass is redistributed, and whether this geometry contains information not captured by realised kurtosis.

\begin{figure}[!htbp]
\centering
\includegraphics[width=0.94\textwidth]
{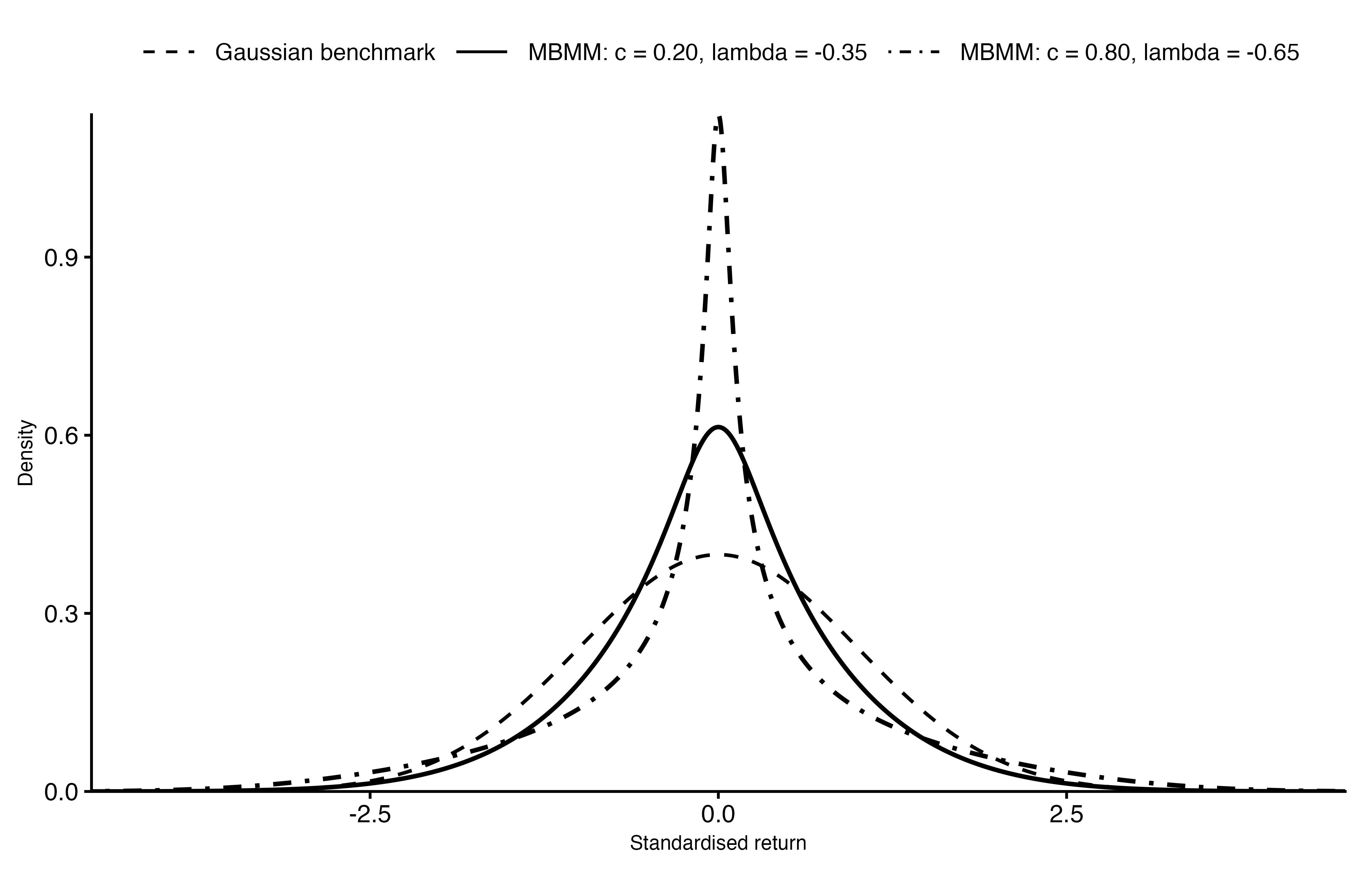}
\caption{
Implied MBMM densities for two combinations of the shape parameter
\(c\) and feedback parameter \(\lambda\), together with the Gaussian
benchmark. The distributions are shown on the standardised scale,
with \(\mu=0\) and \(\sigma=1\), and therefore $\alpha = \lambda$. Different parameter combinations
modify both the location and intensity of the probability-mass
redistribution across the \textcolor{black}{peak}, intermediate returns, and outer
tails.
}
\label{fig:mbmm_density_different_parameters}
\end{figure}

Figure~\ref{fig:mbmm_density_different_parameters} illustrates the
distributional consequences of changing both \(c\) and \(\lambda\).
The two parameter configurations generate visibly different central
concentrations and tail profiles relative to the Gaussian benchmark.
This comparison clearly shows why that the MBMM approach allows better fitting the empirical returns distribution. Moreover, the parameters describe the
entire geometry of the return distribution rather than merely
determining a single measure of kurtosis.

\begin{figure}[!htbp]
\centering
\includegraphics[width=0.92\textwidth]{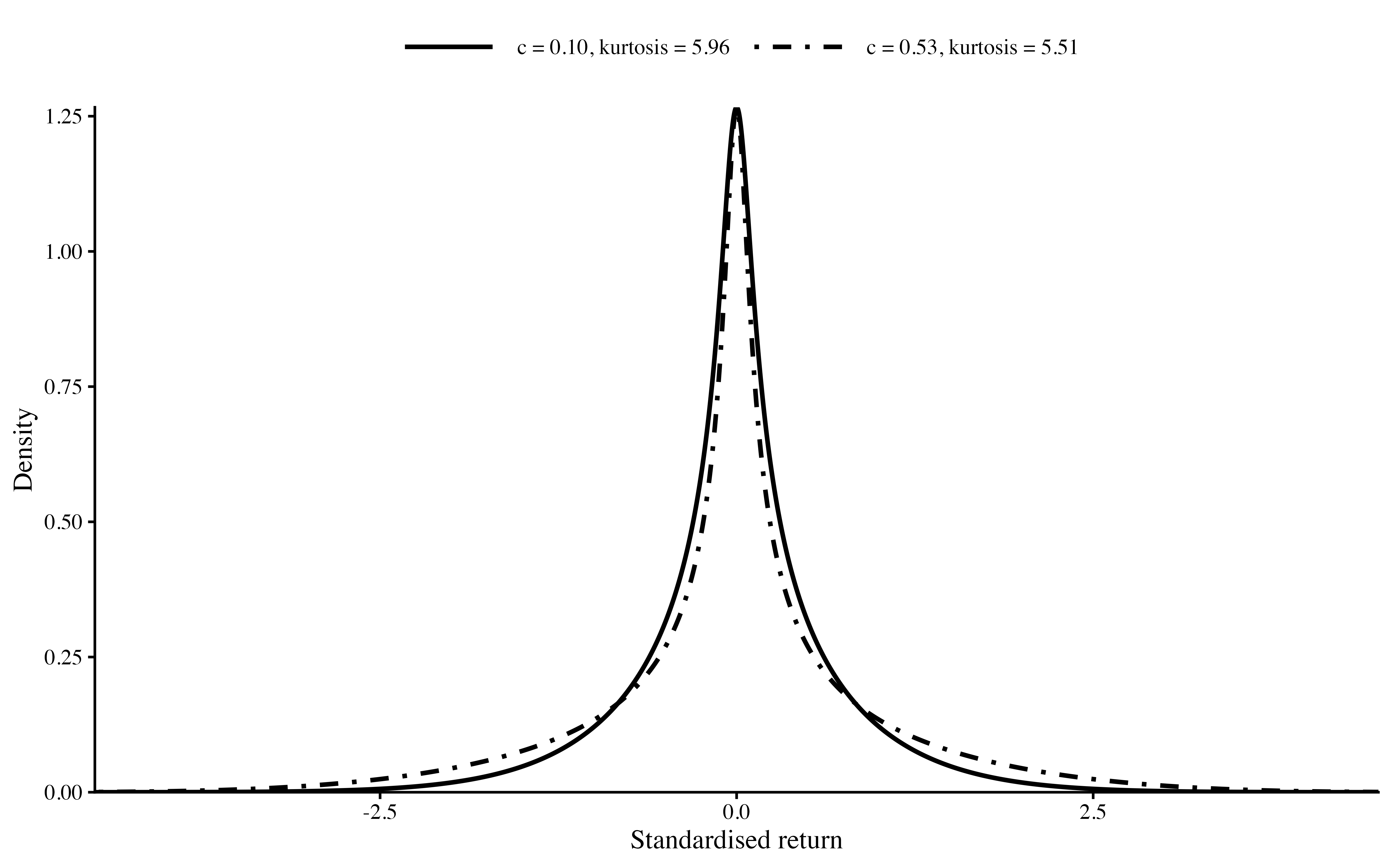}
\caption{
Two MBMM densities obtained by fixing the feedback magnitude at
\(\lambda=-0.685\) and varying only the geometry parameter \(c\).
The two configurations, \(c=0.10\) and \(c=0.53\), imply similar
model-implied kurtosis values, approximately \(5.96\) and \(5.51\),
respectively, despite a more than fivefold change in \(c\). The figure
shows that relatively similar kurtosis values may still correspond to
visibly different return-shape geometries, especially in the central,
shoulder, and tail regions.
}
\label{fig:similar_kurtosis_different_c}
\end{figure}

Figure~\ref{fig:similar_kurtosis_different_c} isolates the role of the
geometry parameter \(c\) by holding the feedback magnitude \(\lambda\)
fixed. 
The two curves correspond to \(c=0.10\) and \(c=0.53\), which
imply model-implied kurtosis values of approximately \(5.96\) and
\(5.51\), respectively. Thus, the difference in kurtosis is relatively
small in comparison with the change in \(c\), which increases by a
factor greater than five. This confirms that \(c\) is not
proportional to kurtosis and, more importantly, that kurtosis alone
does not adequately describe the geometry of the MBMM deformation. In particular, a modest difference in kurtosis does not by itself
reveal how probability mass is redistributed across the \textcolor{black}{peak}, the
intermediate part of the distribution, and the outer tails. Yet
Figure~\ref{fig:similar_kurtosis_different_c} shows that these
geometric differences are already visible even when kurtosis remains
fairly close. 

In other words, one would not infer from kurtosis alone
whether the relevant distinction lies mainly in a sharper central
peak, a more pronounced shoulder depletion, or a different tail
profile. This observation motivates the next step of the methodology:
rather than relying on a single fourth-moment summary, we formally
derive the geometry-induced landmarks that characterise the MBMM
transformation and use them to construct a peak--shoulder--tail
decomposition of the return distribution.

\subsection{Peak--shoulder--tail decomposition}
\label{subsec:peak_shoulder_tail}

Figure~\ref{fig:similar_kurtosis_different_c} shows that two MBMM
specifications can imply fairly similar kurtosis while producing
visibly different return densities, with differences in central
concentration, in the intermediate part of the distribution, and in
the outer tails. To describe these differences, we require a
measurement that records where the fitted return density gains
or loses probability mass relative to the Gaussian benchmark, rather
than a single scalar summary. This decomposition is the primary object
of the paper.

\color{black}
We define the decomposition directly from the difference between the fitted
MBMM density and the Gaussian benchmark. For parameter configurations for
which the transformation
\(
h_{c,\alpha}(z)=z+\alpha f_c(z)
\)
is one-to-one, the density induced on the standardised return scale follows
from the usual change-of-variables formula. Let
\(z=h_{c,\alpha}^{-1}(x)\). Then
\begin{equation}
g_{c,\alpha}(x)
=
\frac{
\phi\!\left(h_{c,\alpha}^{-1}(x)\right)
}{
\left|
1+\alpha
f'_c\!\left(h_{c,\alpha}^{-1}(x)\right)
\right|
}.
\label{eq:mbmm_density}
\end{equation}
Thus, the MBMM density depends both on the displacement of the Gaussian
innovations and on the local compression or expansion induced by the
transformation. This distinction is important because the feedback landmarks
introduced below are defined on the latent innovation scale and therefore need
not coincide with maxima, minima, or crossing points of the resulting return
density.

We then define the density difference
\begin{equation}
\delta(x;c,\alpha)
=
g_{c,\alpha}(x)-\phi(x).
\label{eq:density_difference}
\end{equation}
\color{black}
In the empirically relevant cases considered here, \(\delta(x;c,\alpha)\)
typically follows a peak--shoulder--tail pattern: the MBMM density is
above the Gaussian benchmark near the centre, below it over an
intermediate region, and again above it in the outer tails. We
therefore define the positive crossing points \(x_1(c,\alpha)\) and
\(x_2(c,\alpha)\), with \(0<x_1<x_2\), as the two positive solutions of
\[
\delta(x;c,\alpha)=0.
\]
These crossings induce the following partition of the return scale:
\[
\mathcal P(c,\alpha)=\{x:|x|\le x_1(c,\alpha)\},
\quad
\mathcal S(c,\alpha)=\{x:x_1(c,\alpha)<|x|\le x_2(c,\alpha)\},
\quad
\mathcal T(c,\alpha)=\{x:|x|>x_2(c,\alpha)\}.
\]
We refer to these as the peak, shoulder, and tail regions. The
boundaries \(x_1\) and \(x_2\) are properties of the fitted density on
the return scale: they are the points at which the MBMM density crosses
the Gaussian benchmark, and they depend jointly on \(c\) and
\(\alpha\).

This construction has a direct interpretation. The peak region
contains central excess density relative to the Gaussian benchmark.
The shoulder region contains the corresponding probability-mass
deficit at intermediate returns. The tail region contains excess outer
probability associated with extreme returns. From a financial point of
view, these three regions distinguish between unusually frequent small
movements, a relative scarcity of medium-sized movements, and a higher
incidence of extreme returns.

To summarise these effects, we define the directional area measures
\begin{align}
A_{\mathrm{peak}}(c,\alpha)
&=
\int_{\mathcal P(c,\alpha)}
[\delta(x;c,\alpha)]_+\,dx,
\\
A_{\mathrm{shoulder}}(c,\alpha)
&=
\int_{\mathcal S(c,\alpha)}
[-\delta(x;c,\alpha)]_+\,dx,
\\
A_{\mathrm{tail}}(c,\alpha)
&=
\int_{\mathcal T(c,\alpha)}
[\delta(x;c,\alpha)]_+\,dx,
\end{align}
where \([u]_+=\max(u,0)\). These are not total probabilities of the three regions. Rather, they
measure how much probability mass is reallocated relative to the
Gaussian benchmark. Thus, \(A_{\mathrm{peak}}\) measures central
excess, \(A_{\mathrm{shoulder}}\) measures intermediate-mass
depletion, and \(A_{\mathrm{tail}}\) measures outer-tail excess. Note
that because \(A_{\mathrm{tail}}\) records excess mass beyond \(x_2\),
it can be numerically small even on days with substantial tail risk,
whenever \(x_2\) is located far from the centre. Because both densities integrate to one, any probability mass gained
by the MBMM in some regions must be offset by probability lost
elsewhere. Under the canonical peak--shoulder--tail pattern, this
means that larger values of \(A_{\mathrm{peak}}\) and
\(A_{\mathrm{tail}}\) must be accompanied by a larger shoulder
depletion. Therefore, the three quantities summarise a
redistribution of mass.

\subsection{Feedback landmarks induced by \(c\)}
\label{subsec:shape_landmarks}

The peak--shoulder--tail areas defined above measure how the fitted
return density is deformed relative to the Gaussian benchmark. To
interpret \emph{why} the shape parameter \(c\) produces such
deformations, it is useful to characterise the geometry of the
nonlinear feedback function on the latent innovation scale. We
introduce two landmarks implied directly by \(f_c\). These landmarks
are defined on the latent Gaussian innovation \(z\) and characterise
the mechanism through which the feedback acts on \(Z\), rather than the
crossing points of the resulting return density. The first identifies
where the feedback changes sign; the second identifies where its
absolute magnitude is largest within the central region.

\paragraph{Tail-transition landmark.}

Since \(\arctan(z)\neq0\) for \(z\neq0\), the positive non-zero root of
\(f_c(z)\) is obtained from
\[
2\exp\left(-\frac{cz^2}{2}\right)-1=0.
\]
We define the positive tail-transition landmark as
\begin{equation}
z_T(c)
=
\sqrt{\frac{2\log 2}{c}},
\label{eq:zT}
\end{equation}
with the corresponding symmetric point at \(-z_T(c)\). At
\(z=\pm z_T(c)\), the nonlinear correction vanishes and the
transformation coincides locally with the Gaussian benchmark,
\(h_{c,\alpha}\bigl(\pm z_T(c)\bigr)=\pm z_T(c)\). Under the
orientation \(\alpha<0\), the nonlinear term acts towards the origin
for \(|z|<z_T(c)\) and away from the origin for \(|z|>z_T(c)\); hence
\(z_T(c)\) marks a transition in the direction of the MBMM
deformation, which is why we refer to it as the tail-transition
landmark.

It is important to note that \(z_T(c)\) is a root of the feedback
function on the latent innovation scale. It is \emph{not} the point at
which the tail of the return density begins, nor a crossing point
between the MBMM and Gaussian densities; those crossing points are
\(x_1\) and \(x_2\), defined in
Section~\ref{subsec:peak_shoulder_tail}. The distinction matters
because, for small \(c\), \(z_T(c)\) can lie far in the tails. For
\(c=0.10\),
\[
z_T(c)
=
\sqrt{\frac{2\log 2}{0.10}}
\approx 3.72,
\]
so the feedback does not change from inward to outward action until the
latent innovation is more than \(3.7\) standard deviations from the
origin. This does not imply that the MBMM density differs from the
Gaussian benchmark only beyond that point: the density is affected
throughout its support, because the transformation changes both the
location of the innovations and the Jacobian relating the latent and
observed scales.

\color{black}
\paragraph{Peak-distortion landmark.}
Within the interval \(0<z<z_T(c)\), the factor
\(2\exp(-cz^2/2)-1\) is positive, so that \(f_c(z)>0\), while
\(f_c(0)=f_c(z_T(c))=0\). We therefore define the peak-distortion
landmark as the interior point at which the feedback reaches its largest
magnitude on this central branch,
\begin{equation}
z_P(c)
=
\arg\max_{0<z<z_T(c)} |f_c(z)|
=
\arg\max_{0<z<z_T(c)} f_c(z).
\label{eq:zP}
\end{equation}
Differentiating \(f_c(z)\) gives
\[
f'_c(z)
=
-2cz\exp\left(-\frac{cz^2}{2}\right)\arctan(z)
+
\frac{
2\exp\left(-\frac{cz^2}{2}\right)-1
}{
1+z^2
}.
\]
Hence, \(z_P(c)\) is obtained numerically by selecting the stationary
point in the interval \((0,z_T(c))\), that is, the solution of
\[
2-\exp\left(\frac{cz^2}{2}\right)
-2cz(1+z^2)\arctan(z)
=
0,
\qquad
0<z<z_T(c).
\]
The restriction to \((0,z_T(c))\) is important: \(z_P(c)\) denotes
the inner stationary point, closer to the origin, associated with the
central feedback region, rather than an outer stationary point of
\(f_c\). Under the empirically prevalent orientation \(\alpha<0\),
the same location corresponds to the most negative value of the actual
correction \(\alpha f_c(z)\) on the positive central branch. It is not
the mode of the MBMM density and not necessarily the point at which
the difference between the MBMM and Gaussian densities is largest.
\color{black}

Table~\ref{tab:mbmm_four_landmarks} reports the resulting landmarks for
selected values of \(c\).

\begin{figure}[!htbp]
\centering
\includegraphics[width=\textwidth]
{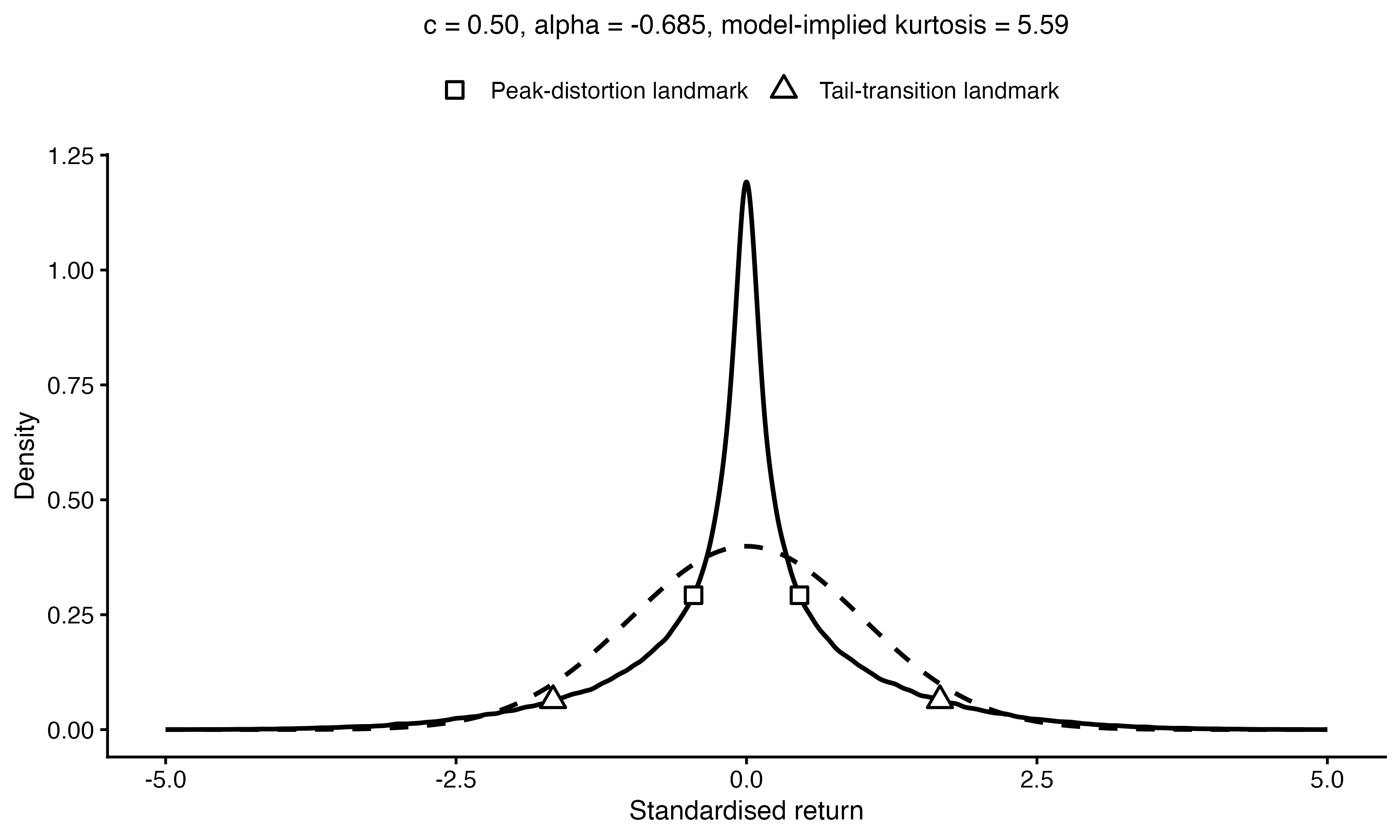}
\caption{
\textcolor{black}{MBMM density for \(c=0.50\) and \(\alpha=-0.685\), together with the
Gaussian benchmark. Squares indicate the return-scale images of the
peak-distortion landmarks,
\(x_P=h_{c,\alpha}(z_P)\), whereas triangles indicate the
return-scale images of the tail-transition landmarks,
\(x_T=h_{c,\alpha}(z_T)\). The peak-distortion landmarks identify the
locations associated with the strongest central feedback correction,
while the tail-transition landmarks identify where the feedback
changes from inward to outward action. The latter should not be
interpreted as the empirical onset of the density tails.
}}
\label{fig:single_density_landmarks}
\end{figure}

\begin{table}[!htbp]
\centering
\small
\setlength{\tabcolsep}{7pt}
\renewcommand{\arraystretch}{1.12}
\caption{
MBMM peak-distortion and tail-transition landmarks for selected values
of \(c\). The peak-distortion points are obtained numerically from the
first-order condition following Equation~\eqref{eq:zP}, restricting
the solution to \(0<z<z_T(c)\).
}
\label{tab:mbmm_four_landmarks}
\begin{tabular}{rrrrr}
\toprule
\(c\)
&
\(-z_T(c)\)
&
\(-z_P(c)\)
&
\(z_P(c)\)
&
\(z_T(c)\)
\\
\midrule
0.1 & -3.7233 & -1.4602 & 1.4602 & 3.7233 \\
0.2 & -2.6328 & -1.1269 & 1.1269 & 2.6328 \\
0.3 & -2.1496 & -0.9619 & 0.9619 & 2.1496 \\
0.4 & -1.8616 & -0.8567 & 0.8567 & 1.8616 \\
0.5 & -1.6651 & -0.7815 & 0.7815 & 1.6651 \\
0.6 & -1.5200 & -0.7240 & 0.7240 & 1.5200 \\
0.7 & -1.4073 & -0.6780 & 0.6780 & 1.4073 \\
0.8 & -1.3164 & -0.6401 & 0.6401 & 1.3164 \\
0.9 & -1.2411 & -0.6080 & 0.6080 & 1.2411 \\
1.0 & -1.1774 & -0.5805 & 0.5805 & 1.1774 \\
\bottomrule
\end{tabular}
\end{table}
Figure~\ref{fig:single_density_landmarks} displays the two landmarks
for one representative MBMM specification.
Squares indicate the return-scale images of the peak-distortion points,
while triangles indicate the return-scale images of the tail-transition
points. The squares lie in the intermediate part of the density because
they mark where the inward feedback is strongest, not where the density
reaches its maximum. The triangles lie much farther out because, for
small values of \(c\), the sign change in the feedback function occurs
only for relatively large Gaussian innovations.

The landmarks therefore provide an interpretive device that connects
the shape parameter \(c\) to the peak--shoulder--tail geometry of the
previous subsection. A smaller \(c\) implies a larger \(z_T(c)\),
placing the feedback transition farther from the centre and producing a
wide-transition geometry; a larger \(c\) moves the transition towards
the origin and compresses the geometry. This link is used in the
empirical analysis to interpret why stock-days with similar realised
kurtosis can display different fitted shapes.

\section{Simulation-based minimum-distance estimation}
\label{sec:estimation}

\textcolor{black}{Previous applications of the MBMM adopt a predominantly simulation-based calibration approach to estimate the parameters governing the nonlinear
feedback. In particular, \cite{DhesiShakeelXiao2016} represent the empirical return
distribution through a histogram, construct customised bins, and obtain the
corresponding model-implied frequencies by repeatedly simulating the MBMM.
Candidate values of the feedback parameters are then selected according to
the resulting $\chi^2$ goodness-of-fit criterion. Although useful in practice, this procedure remains relatively
heuristic from an estimation perspective. Indeed, the resulting fit
depends on the histogram construction and on the associated bin boundaries,
while parameter selection is performed by exploring candidate parameter
configurations rather than by defining and directly optimising a continuous
sample criterion. These features are especially inconvenient in our setting,
where the model must be estimated repeatedly from relatively short intraday
samples for a large number of stock-days.}

\textcolor{black}{We therefore adopt a different
estimation perspective, borrowing the same minimum-distance philosophy. Instead of
matching binned frequencies, we match empirical and model-implied quantiles
and estimate the MBMM parameters by directly minimising their quadratic
distance. The resulting simulation-based minimum-distance estimator preserves
the simulation logic of the original MBMM calibration, while providing a
bin-free and explicitly defined optimisation criterion that can be applied
uniformly across stock-days. Let us now discuss the procedure adopted estimation procedure more in details} 

Let
\(
\mathcal R_t
=
\{r_{t,1},\ldots,r_{t,n_t}\}
\)
denote the asset returns observed on time period \(t\). We replace the mean and standard deviation with their
empirical counterparts, that is,
\[
\widehat\mu_t
=
\frac{1}{n_t}\sum_{j=1}^{n_t}r_{t,j},
\]
and
\[\widehat\sigma_t^2
=
\frac{1}{n_t-1}
\sum_{j=1}^{n_t}
(r_{t,j}-\widehat\mu_t)^2.
\]
Conditional on these quantities, the parameters of interest are
\[
\vartheta=(c,\lambda)^\top,
\qquad
c>0,
\quad
\lambda\in\mathbb R.
\]
The estimation principle is to select the parameter values whose
model-implied return distribution most closely reproduces the
empirical distribution observed on day \(t\). Since the MBMM
distribution is generated through a nonlinear transformation of a
Gaussian innovation, its quantiles are approximated by simulation. Let
\[
\mathcal T_Q
=
\{\tau_1,\ldots,\tau_Q\},
\qquad
0<\tau_1<\cdots<\tau_Q<1,
\]
be a fixed grid of quantile probabilities, and let
\(\widehat q_t(\tau_q)\) denote the empirical
\(\tau_q\)-quantile of the observed returns. For a candidate parameter
vector \(\vartheta=(c,\lambda)^\top\), we generate \(B\) artificial
samples of the same length as the observed sample:
\begin{equation}
r^{\ast}_{t,b,j}(\vartheta)
=
\widehat\mu_t
+
\widehat\sigma_t Z_{t,b,j}
+
\lambda f_c(Z_{t,b,j}),
\qquad
Z_{t,b,j}\overset{\mathrm{iid}}{\sim}\mathcal N(0,1).
\label{eq:simulated_returns_estimation}
\end{equation}
For each simulated sample, let
\(\widehat q^{\ast}_{t,b}(\tau_q;\vartheta)\) denote the corresponding
sample quantile. The simulation-averaged model quantile is
\[
\overline q^{\ast}_{t,B}(\tau_q;\vartheta)
=
\frac{1}{B}
\sum_{b=1}^{B}
\widehat q^{\ast}_{t,b}(\tau_q;\vartheta).
\]
Collecting empirical and simulated quantiles in the vectors
\[
\widehat{\mathbf q}_t
=
\bigl(
\widehat q_t(\tau_1),\ldots,\widehat q_t(\tau_Q)
\bigr)^\top
\]
and
\[
\overline{\mathbf q}^{\ast}_{t,B}(\vartheta)
=
\bigl(
\overline q^{\ast}_{t,B}(\tau_1;\vartheta),
\ldots,
\overline q^{\ast}_{t,B}(\tau_Q;\vartheta)
\bigr)^\top,
\]
the minimum-distance criterion is
\begin{equation}
Q_{t,B}(\vartheta)
=
\left[
\widehat{\mathbf q}_t
-
\overline{\mathbf q}^{\ast}_{t,B}(\vartheta)
\right]^\top
W_Q
\left[
\widehat{\mathbf q}_t
-
\overline{\mathbf q}^{\ast}_{t,B}(\vartheta)
\right],
\label{eq:minimum_distance_criterion}
\end{equation}
where \(W_Q\) is a positive-definite weighting matrix. The baseline
implementation uses \(W_Q=I_Q\), so that all selected quantiles receive
equal weight. The daily estimator is
\begin{equation}
\widehat\vartheta_t
=
\begin{pmatrix}
\widehat c_t\\
\widehat\lambda_t
\end{pmatrix}
=
\arg\min_{\vartheta\in\Theta}
Q_{t,B}(\vartheta).
\label{eq:smde_estimator}
\end{equation}
The estimator therefore searches for the feedback geometry that makes
the simulated distribution as close as possible to the empirical
distribution. Matching a grid of quantiles uses information from the
centre, shoulders, and tails simultaneously. This is materially
different from matching a single statistic such as variance or
kurtosis, which would reproduce only one feature of the observed
distribution.

At the population level, the criterion is minimised when the
model-implied quantile function coincides with the quantile function of
the data-generating distribution over the selected probability grid.
As the sample size, the number of simulations, and the resolution of
the quantile grid increase, the sample criterion provides an
increasingly accurate approximation to this population distance. In
one dimension, the equally weighted criterion can also be interpreted
as a discretised quadratic Wasserstein distance between the empirical
and model-implied return distributions. To reduce simulation noise during numerical optimisation, the same
Gaussian draws are retained across candidate parameter values within
each estimation. These common random numbers make the objective
function smoother and ensure that changes in the criterion primarily
reflect changes in \((c,\lambda)\), rather than new simulation draws.

We finally remark that the use of the full return distribution is important not only for
estimation accuracy, but also for identification. Indeed, a procedure based
only on matching kurtosis would select parameter values satisfying
approximately
\(
\kappa(c,\lambda)
=
\kappa_{\mathrm{data}}
\), which is one scalar restriction for two unknown shape parameters. Consequently, several combinations of \(c\) and \(\lambda\) may
produce the same, or nearly the same, kurtosis while generating
different return densities. This feature was illustrated in
Figure~\ref{fig:similar_kurtosis_different_c}: substantially different
values of \(c\), at the same value of \(\lambda\), produced similar
kurtosis but visibly different central and intermediate-return
geometries. Matching kurtosis alone would treat these specifications as
nearly equivalent, despite their different peak, shoulder, and tail
structures. Hence, kurtosis is useful as a descriptive summary, but it
is not an identifying statistic for the two MBMM shape parameters. The quantile-based criterion instead compares a collection of
distributional features. Central quantiles contain information on peak
concentration, intermediate quantiles reflect the shoulder region, and
extreme quantiles capture tail behaviour. The joint pattern of these
quantiles distinguishes parameter combinations that cannot be
separated using kurtosis alone. The case \(\lambda=0\) is fundamentally different. When the nonlinear
coefficient is zero,
\(
r_t
=
\mu+\sigma Z_t
\)
regardless of \(c\). Therefore, \(c\) has no effect on the observable
distribution and is not identified under the Gaussian benchmark. A numerical optimiser may nevertheless return a particular estimate
of \(c\), possibly a very large or very small value. Such a value is
only an artefact of numerical search over a flat objective and has no
statistical interpretation. More generally, when
\(|\widehat\lambda|\) is close to zero, the objective is nearly flat in
the \(c\)-direction, so that \(c\) may be weakly identified and
unstable in finite samples. For this reason, estimates of \(c\) are
interpreted only when the fitted nonlinear component is meaningfully
different from zero.

\color{black}
\section{Monte Carlo simulation study}
\label{sec:simulation}

We conduct a Monte Carlo experiment to examine the finite-sample behaviour of the
simulation-based quantile minimum-distance estimator. The experiment addresses two
complementary questions. First, when the data-generating process is MBMM, how accurately
does the estimator recover the geometry parameter \(c\) and the nonlinear distortion
parameters \(\lambda\) and \(\alpha=\lambda/\sigma\)? Second, when the data are generated
by the Gaussian benchmark, does the estimated nonlinear component concentrate around
zero rather than introducing spurious MBMM structure?

\paragraph{DGP 1: MBMM}

Under the nonlinear data-generating process,
\begin{equation}
r_t
=
\mu_0+\sigma_0 Z_t+\lambda_0 f_{c_0}(Z_t),
\qquad
Z_t\overset{\mathrm{iid}}{\sim}\mathcal N(0,1),
\label{eq:simulation_mbmm}
\end{equation}
with
\[
\mu_0=0.0002,
\qquad
\sigma_0=0.01,
\qquad
\lambda_0=\alpha_0\sigma_0.
\]
For the sake of robustness, we consider four
distinct MBMM designs obtained by combining two levels of distortion intensity with
positive and negative distortions:
\[
(c_0,\alpha_0)
\in
\left\{
(0.6,0.4),
(0.6,-0.4),
(0.8,0.6),
(0.8,-0.6)
\right\}.
\]
The corresponding values of \(\lambda_0\) are \(0.004\), \(-0.004\), \(0.006\),
and \(-0.006\), respectively. The designs therefore cover moderate and stronger
nonlinear deformations while allowing the orientation of the feedback effect to change.

\paragraph{DGP 2: Gaussian benchmark}

Under the Gaussian benchmark,
\begin{equation}
r_t
=
\mu_0+\sigma_0 Z_t,
\qquad
Z_t\overset{\mathrm{iid}}{\sim}\mathcal N(0,1),
\label{eq:simulation_gbm}
\end{equation}
which corresponds to \(\lambda_0=\alpha_0=0\). In this case the nonlinear component
vanishes and \(c\) is not identified. Consequently, no bias or RMSE for
\(\widehat c\) is reported under the Gaussian DGP; the relevant question is instead
whether \(\widehat\lambda\) and \(\widehat\alpha\) concentrate around zero.

For both DGPs we consider \(T\in\{100,500,1000\}\). Each MBMM configuration is based
on \(M=300\) independent Monte Carlo replications and, within each replication, estimation
uses \(B=100\) simulated samples.
Common Gaussian draws are retained throughout each numerical optimisation in order to
reduce simulation noise. 
The estimator performance is evaluated through bias and root mean squared error,
\begin{equation}
\operatorname{Bias}(\widehat\theta)
=
\frac{1}{M}
\sum_{m=1}^{M}
(\widehat\theta_m-\theta_0),
\qquad
\operatorname{RMSE}(\widehat\theta)
=
\left[
\frac{1}{M}
\sum_{m=1}^{M}
(\widehat\theta_m-\theta_0)^2
\right]^{1/2}.
\label{eq:simulation_performance}
\end{equation}
Because the four design labels correspond to the same DGP when
\(\lambda_0=\alpha_0=0\), the Gaussian results in Table~\ref{tab:simulation_results}
pool the four equivalent simulation blocks, yielding \(1{,}200\) replications for each
value of \(T\).

\begin{table}[!htbp]
\centering
\caption{Monte Carlo performance of the simulation-based quantile minimum-distance estimator}
\label{tab:simulation_results}
\scriptsize
\setlength{\tabcolsep}{4.2pt}
\renewcommand{\arraystretch}{1.10}
\resizebox{\textwidth}{!}{%
\begin{tabular}{llrrrrrr}
\toprule
Design
& \(T\)
& Bias(\(\widehat c\))
& RMSE(\(\widehat c\))
& Bias(\(\widehat\lambda\))
& RMSE(\(\widehat\lambda\))
& Bias(\(\widehat\alpha\))
& RMSE(\(\widehat\alpha\)) \\
\midrule
\multicolumn{8}{l}{\textit{Panel A: MBMM DGP}} \\
\addlinespace[2pt]

Moderate positive
& 100
& 0.0146
& 0.1642
& -0.000011
& 0.001312
& -0.0151
& 0.1225 \\

& 500
& 0.0146
& 0.0148
& 0.000034
& 0.000760
& -0.0089
& 0.0736 \\

& 1000
& 0.0138
& 0.0152
& 0.000131
& 0.000790
& 0.0002
& 0.0750 \\
\addlinespace

Moderate negative
& 100
& 0.1075
& 0.1926
& -0.000627
& 0.004604
& -0.0762
& 0.4708 \\

& 500
& 0.0700
& 0.0822
& -0.000130
& 0.000980
& -0.0194
& 0.1010 \\

& 1000
& 0.0682
& 0.0798
& -0.000190
& 0.001645
& -0.0276
& 0.1709 \\
\addlinespace

Strong positive
& 100
& -0.0161
& 0.1384
& -0.000133
& 0.001223
& -0.0091
& 0.1166 \\

& 500
& -0.0075
& 0.0211
& -0.000077
& 0.000818
& -0.0074
& 0.0806 \\

& 1000
& -0.0081
& 0.0160
& -0.000054
& 0.000738
& -0.0037
& 0.0738 \\
\addlinespace

Strong negative
& 100
& -0.0122
& 0.1503
& -0.000355
& 0.003701
& -0.0120
& 0.3607 \\

& 500
& -0.0143
& 0.0847
& 0.000101
& 0.001053
& 0.0354
& 0.1066 \\

& 1000
& -0.0141
& 0.0143
& 0.000308
& 0.000902
& 0.0544
& 0.0988 \\

\midrule
\multicolumn{8}{l}{\textit{Panel B: Gaussian benchmark}} \\
\addlinespace[2pt]

GBM
& 100
& --
& --
& 0.000002
& 0.002321
& -0.0007
& 0.2431 \\

& 500
& --
& --
& 0.000017
& 0.000926
& 0.0017
& 0.0928 \\

& 1000
& --
& --
& -0.000011
& 0.000854
& -0.0011
& 0.0853 \\
\bottomrule
\end{tabular}%
}

\vspace{0.15cm}

\begin{minipage}{0.94\textwidth}
\footnotesize
\textit{Notes:}
For the moderate designs, \(c_0=0.6\) and \(|\alpha_0|=0.4\), implying
\(|\lambda_0|=0.004\). For the strong designs, \(c_0=0.8\) and
\(|\alpha_0|=0.6\), implying \(|\lambda_0|=0.006\). Positive and negative
labels refer to the sign of \(\alpha_0\) and \(\lambda_0\). Each MBMM
configuration uses \(M=300\) Monte Carlo replications, \(B=100\) simulated
samples within the objective, and \(Q=51\) quantile probabilities. Under the
Gaussian benchmark, \(\lambda_0=\alpha_0=0\) and \(c\) is not identified.
The GBM rows pool the four equivalent Gaussian simulation blocks, for a total
of \(1{,}200\) replications at each \(T\).
\end{minipage}
\end{table}

Table~\ref{tab:simulation_results} provides interesting finite-sample
evidence for the proposed estimator across both positive and negative
distortion configurations. In the positive-feedback designs, the parameter
recovery is particularly accurate. For the strong positive configuration,
the bias of $\widehat c$ is already small at $T=100$ and becomes negligible
for $T=500$ and $T=1000$, while the RMSE falls sharply as the sample size
increases. The distortion parameter $\alpha$ is also recovered accurately,
with bias close to zero and steadily declining RMSE. A similar pattern emerges in the moderate positive design. The bias of
$\widehat c$ remains close to zero across all sample sizes, while its RMSE
becomes very small once $T$ reaches 500 observations. The estimator also
recovers $\alpha$ accurately, with virtually no bias at $T=1000$. The negative-feedback configurations are somewhat more difficult in small
samples, particularly for the distortion intensity. Nevertheless, estimator
accuracy improves substantially as the sample size increases. In the
moderate negative design, the RMSE of $\widehat c$ decreases from $0.1926$
at $T=100$ to approximately $0.08$ for $T=500$ and $T=1000$. The RMSE of
$\widehat\alpha$ also falls sharply between $T=100$ and $T=500$, although
some non-monotonicity remains at the largest sample size. In the strong
negative design, both parameters display small bias, while their RMSEs
decrease markedly with $T$. In particular, the RMSE of $\widehat c$ falls
to $0.0143$ at $T=1000$, and that of $\widehat\alpha$ falls below $0.10$. The Gaussian benchmark confirms that the estimator does not systematically
introduce spurious nonlinear structure when the true data-generating process
is Gaussian. Since $\lambda_0=\alpha_0=0$, the parameter $c$ is not
identified and is therefore not evaluated. The bias of $\widehat\lambda$
remains essentially zero at all sample sizes, while the RMSE of both
$\widehat\lambda$ and $\widehat\alpha$ declines substantially as $T$
increases.

Therefore, the simulation results indicate that the proposed
simulation-based quantile minimum-distance estimator performs well over a
range of MBMM configurations. These results also support the use of the proposed procedure for repeated intraday estimation in the empirical analysis that follows.

\color{black}
\section{Empirical analysis}
\label{sec:empirical_analysis}

\subsection{Data}
\label{sec:data}

The empirical analysis is conducted at the stock-day level using
one-minute intraday observations for ten highly liquid U.S. equities:
AAPL, AMZN, GOOGL, JPM, META, MSFT, NVDA, TSLA, UNH, and XOM.
The data are obtained from the Alpaca Market\footnote{Data can be retrived by API at \url{https://alpaca.markets/data}.} and cover the period from 3 May 2021 to
30 April 2026. All timestamps are converted from UTC to New York time, and only
observations recorded during the regular trading session, from 09:30 to 16:00, are retained. Returns are computed exclusively within each trading day, so overnight price changes are excluded. A complete
regular-session stock-day contains \(389\) one-minute log returns, and
only complete stock-days are retained in the final estimation sample. Let \(P_{i,t,j}\) denote the price of stock \(i\) on trading day \(t\)
at intraday minute \(j\). The corresponding one-minute log return is
defined as
\begin{equation}
r_{i,t,j}
=
\log P_{i,t,j}
-
\log P_{i,t,j-1},
\qquad
j=1,\ldots,389.
\label{eq:data_intraday_returns}
\end{equation}
Hence, for every stock-day, the empirical object of interest is not a
single daily return but an entire within-day distribution containing
389 high-frequency observations.

\begin{figure}[!htbp]
\centering
\includegraphics[width=\textwidth]
{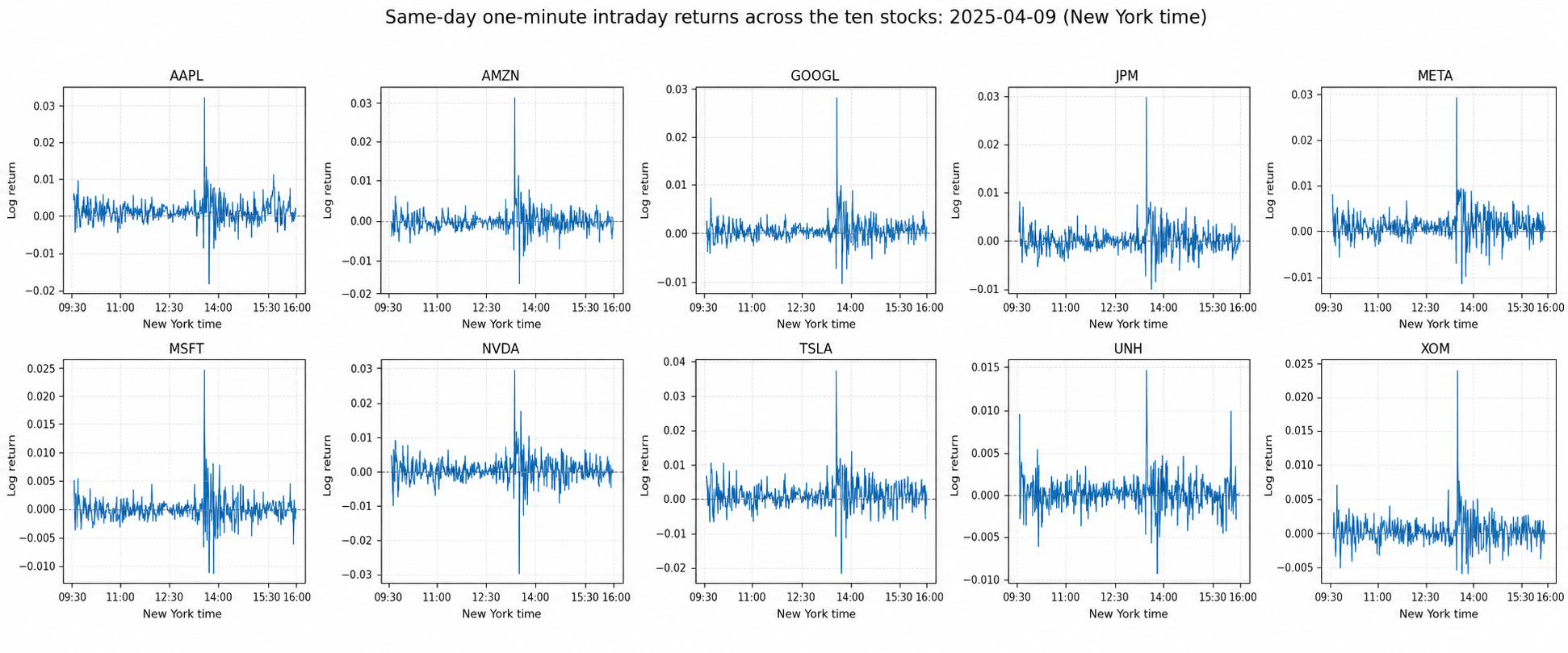}
\caption{
Same-day one-minute intraday log returns for the ten stocks on
9 April 2025. Returns are computed exclusively within the regular
09:30--16:00 New York trading session. The figure provides an
illustration of the high-frequency observations used to construct the
daily empirical return distributions analysed in the paper.
}
\label{fig:same_day_intraday_returns}
\end{figure}

\begin{figure}[!htbp]
\centering
\includegraphics[width=\textwidth]
{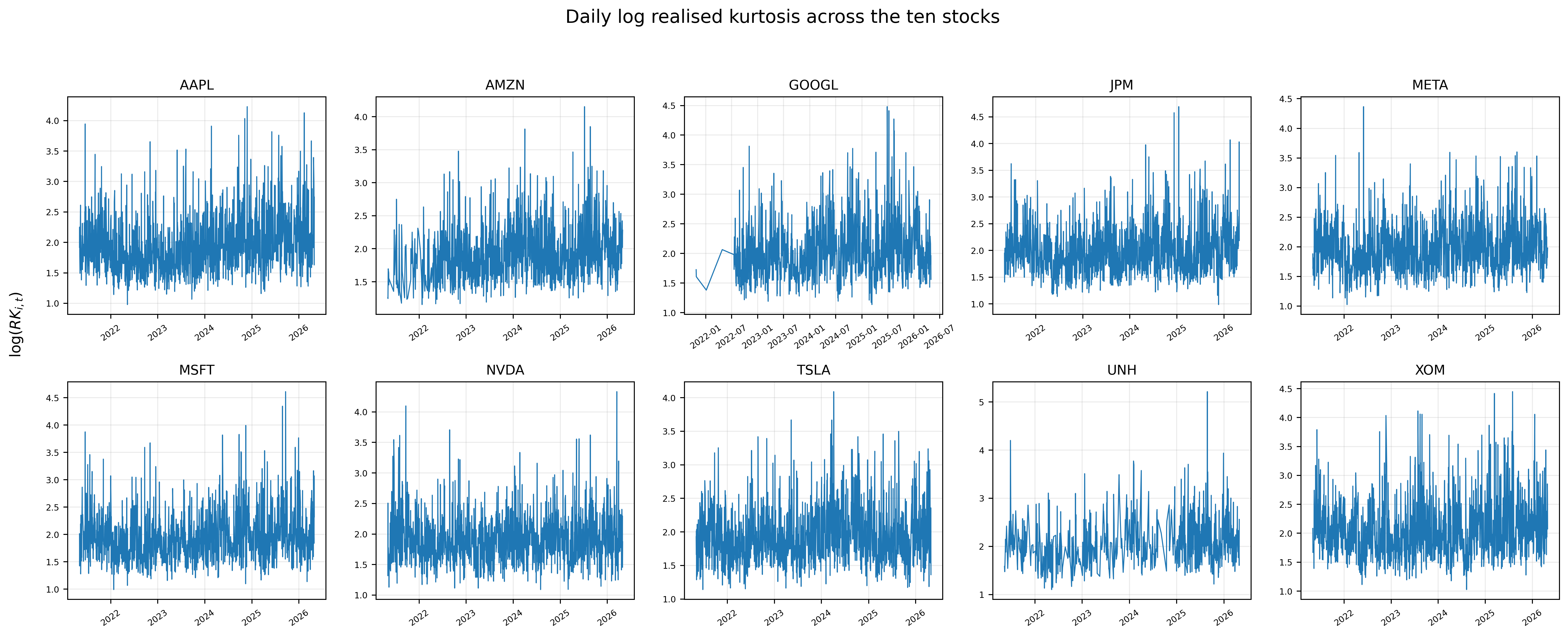}
\caption{
Daily log realised kurtosis for the ten stocks over the sample period.
For each stock-day, realised kurtosis is computed from the 389
regular-session one-minute log returns. The logarithmic scale is used
to improve graphical readability and highlights the substantial
time variation and recurrent spikes in intraday tail thickness.
}
\label{fig:panel_logrk}
\end{figure}

Figure~\ref{fig:same_day_intraday_returns} provides an illustration of
the raw data by reporting the one-minute return paths of the ten stocks
on 9 April 2025. 
For each stock-day, we compute the intraday sample mean
\(\widehat\mu_{i,t}\), the intraday standard deviation
\(\widehat\sigma_{i,t}\), and realised kurtosis. The latter is defined
as the Pearson fourth standardised moment:
\begin{equation}
RK_{i,t}
=
\frac{
389^{-1}
\sum_{j=1}^{389}
\left(
r_{i,t,j}-\widehat\mu_{i,t}
\right)^4
}{
\left[
389^{-1}
\sum_{j=1}^{389}
\left(
r_{i,t,j}-\widehat\mu_{i,t}
\right)^2
\right]^2
}.
\label{eq:data_realised_kurtosis}
\end{equation}
The quantity \(RK_{i,t}\) therefore provides a daily measure of the
overall tail thickness of the one-minute return distribution. A
Gaussian intraday distribution has population kurtosis equal to three,
whereas larger values indicate a greater contribution of extreme
observations to the fourth standardised moment. Figure~\ref{fig:panel_logrk} reports the daily evolution of
\(\log(RK_{i,t})\) for all ten stocks over the sample period. 
The
logarithmic transformation is used only for graphical readability,
since realised kurtosis is strongly right-skewed and occasionally
takes very large values. The panels show substantial variation over
time, with recurrent spikes, periods of relative stability, and
stock-specific episodes of elevated intraday tail thickness. Thus,
the shape of the one-minute return distribution is not constant across
trading days or across stocks. Realised kurtosis, however, provides only one scalar summary of each
daily distribution. It measures the overall magnitude of the fourth
standardised moment, but it does not indicate how the corresponding
distributional shape is generated. Two stock-days may exhibit similar
realised kurtosis while differing substantially in central
concentration, intermediate-return depletion, tail transition, or
outer-tail behaviour. For this reason, the MBMM is estimated separately for every
stock-day. Conditional on the observed intraday location and scale,
the estimation procedure produces daily values of
\(\widehat c_{i,t}\) and \(\widehat\lambda_{i,t}\). Accordingly, each observed value of
\(RK_{i,t}\) is accompanied by a potentially different fitted pair
\(
\left(
\widehat c_{i,t},
\widehat\lambda_{i,t}
\right),
\)
which provides a richer description of the daily intraday return
distribution.

The empirical analysis that follows is organised around three
questions. First, we examine whether the fitted MBMM provides a
meaningfully better in-sample representation of the one-minute return
distribution than the Gaussian GBM benchmark. Second, we investigate
how the estimated parameters \(c\) and \(\lambda\), and the associated
peak--shoulder--tail geometry, vary across stocks and over time,
including cases in which similar realised kurtosis is associated with
different fitted distributional shapes. Third, we study whether these daily MBMM state variables are forecastable. These questions are addressed in the next subsections.

\subsection{In-sample distributional fit}
\label{sec:insample_fit}

Before treating the estimated MBMM parameters as empirical state
variables, we first ask whether the MBMM provides a better in-sample
representation of one-minute returns than the Gaussian GBM benchmark. For each stock-day, let \(Q^{\mathrm{GBM}}_{i,t}\) denote the
quantile-distance objective under the Gaussian benchmark, and let
\(Q^{\mathrm{MBMM}}_{i,t}\) denote the minimised MBMM objective. The
percentage reduction in the objective is
\begin{equation}
\Delta^{\mathrm{obj}}_{i,t}
=
100
\frac{
Q^{\mathrm{GBM}}_{i,t}
-
Q^{\mathrm{MBMM}}_{i,t}
}{
Q^{\mathrm{GBM}}_{i,t}
}.
\label{eq:insample_objective_improvement}
\end{equation}
Table~\ref{tab:one_minute_measurement_quality} shows that the answer
is clearly yes. 
\begin{table}[!htbp]
\centering
\caption{One-minute MBMM distributional fit by stock}
\label{tab:one_minute_measurement_quality}
\small
\setlength{\tabcolsep}{5.8pt}
\renewcommand{\arraystretch}{1.15}
\begin{tabular*}{\textwidth}{@{\extracolsep{\fill}}lrrrrr@{}}
\toprule
Stock
& Mean \(RK\)
& Median \(RK\)
& Mean \(\Delta^{\mathrm{obj}}\)
& Median \(\Delta^{\mathrm{obj}}\)
& Positive days \\
&
&
&
\((\%)\)
&
\((\%)\)
&
\((\%)\) \\
\midrule
AAPL  & 8.023 & 6.356 & 61.61 & 63.76 & 100.00 \\
AMZN  & 7.542 & 6.432 & 63.83 & 66.48 & 100.00 \\
GOOGL & 8.811 & 6.946 & 64.91 & 66.32 & 100.00 \\
JPM   & 8.370 & 6.756 & 62.80 & 64.85 & 100.00 \\
META  & 7.918 & 6.691 & 65.11 & 67.84 & 100.00 \\
MSFT  & 8.000 & 6.548 & 64.06 & 65.97 & 100.00 \\
NVDA  & 7.419 & 6.312 & 55.77 & 58.64 & 100.00 \\
TSLA  & 7.733 & 6.503 & 62.86 & 65.29 & 100.00 \\
UNH   & 9.233 & 7.364 & 68.13 & 70.27 & 100.00 \\
XOM   & 9.094 & 7.110 & 61.40 & 63.40 & 100.00 \\
\midrule
\textbf{Average}
& \textbf{8.214}
& \textbf{6.702}
& \textbf{63.05}
& \textbf{65.28}
& \textbf{100.00} \\
\bottomrule
\end{tabular*}
\end{table}
Across the ten stocks, the mean reduction in the
quantile-distance objective is \(63.05\%\), while the median reduction
is \(65.28\%\). The improvement is positive on every estimator-ready
stock-day. Even the weakest average improvement, observed for NVDA, is
\(55.77\%\), meaning that the MBMM reduces the fitting objective by
more than one half relative to GBM. The strongest average improvement
is observed for UNH, with a mean reduction of \(68.13\%\).

\begin{figure}[!htbp]
\centering
\includegraphics[width=\textwidth]
{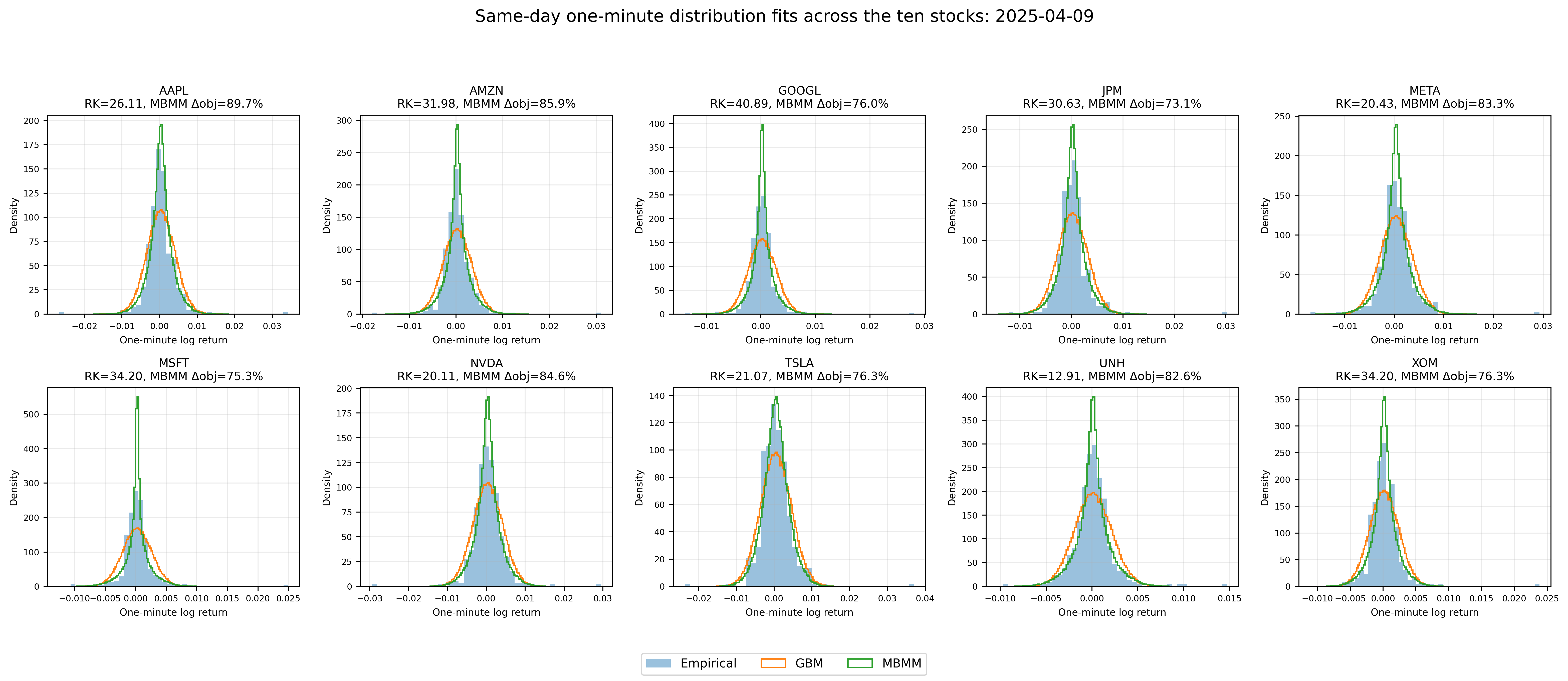}
\caption{
Same-day one-minute distribution fits for the ten stocks on
9 April 2025. The histogram represents the empirical intraday return
distribution, the benchmark curve is the Gaussian GBM fit, and the
alternative curve is the fitted MBMM density. The titles report
realised kurtosis and the percentage reduction in the quantile-distance
objective delivered by MBMM.
}
\label{fig:same_day_distribution_fits}
\end{figure}

Figure~\ref{fig:same_day_distribution_fits} provides a same-day visual
check for 9 April 2025. 
The comparison is useful because the ten stocks
are observed under the same market date, but display different
empirical return shapes. In all panels, the Gaussian benchmark fails to
capture the sharp central concentration of one-minute returns and gives
a poorer description of the intermediate and tail regions. The MBMM
density instead follows the empirical peak and the surrounding
shoulder region more closely. On this day, the objective reduction
ranges from \(73.1\%\) for JPM to \(89.7\%\) for AAPL, with an average
improvement of about \(80.3\%\) across the ten stocks.

\begin{figure}[!htbp]
\centering
\includegraphics[width=\textwidth]
{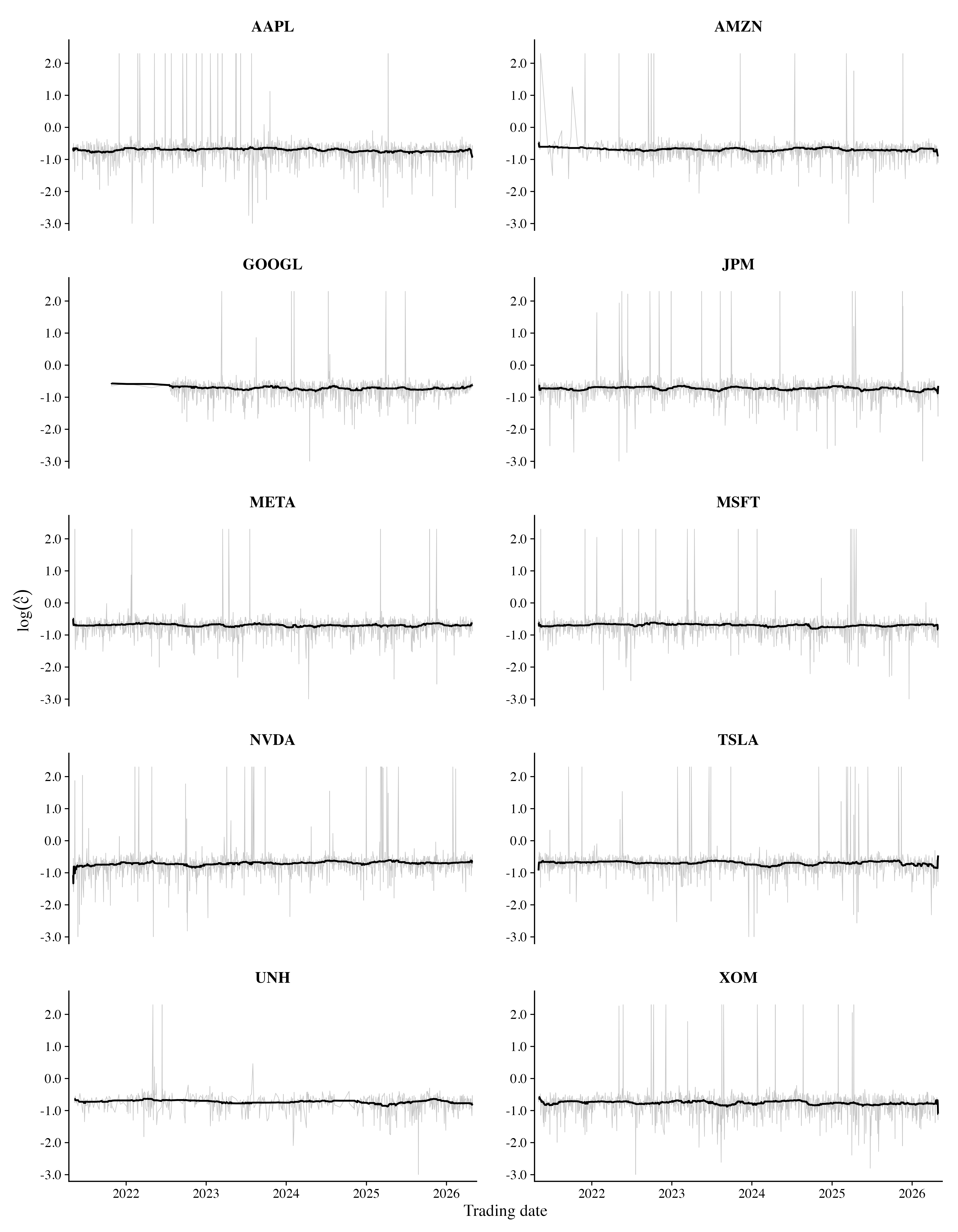}
\caption{
Daily time series of the MBMM shape component,
\(\log(\hat c_{i,t})\), for the ten stocks. The light grey line reports
the daily estimate, while the black line reports a centred rolling
median smoother computed over 63 trading days. The transformation is
used to improve graphical readability and to reduce the visual
dominance of occasional large fitted values of \(\hat c_{i,t}\).
}
\label{fig:smoothed_log_c_timeseries}
\end{figure}

\begin{figure}[!htbp]
\centering
\includegraphics[width=\textwidth]
{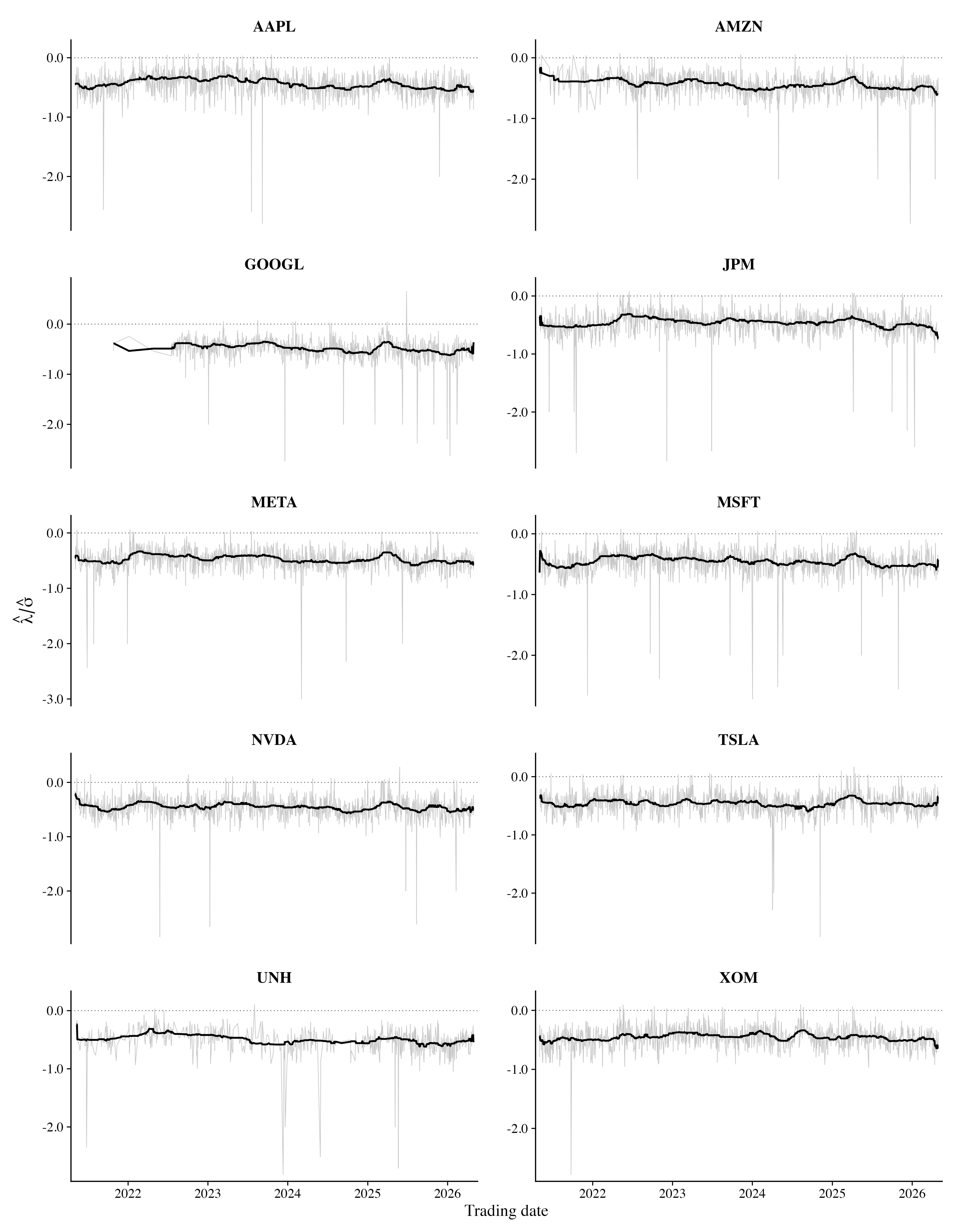}
\caption{
Daily time series of the scale-normalised MBMM distortion intensity,
\(\hat\lambda_{i,t}/\hat\sigma_{i,t}\), for the ten stocks. The light
grey line reports the daily estimate, while the black line reports a
centred rolling median smoother computed over 63 trading days. The
horizontal dotted line marks zero distortion relative to the Gaussian
benchmark.
}
\label{fig:smoothed_alpha_timeseries}
\end{figure}

Figures~\ref{fig:smoothed_log_c_timeseries} and
\ref{fig:smoothed_alpha_timeseries} show that the fitted MBMM
parameters are not purely isolated daily fitting outcomes. 

The smoothed
\(\log(\hat c_{i,t})\) series are relatively stable within each stock,
but display persistent low-frequency movements and occasional abrupt
departures. This indicates that the location of the MBMM shape
transition changes over time rather than remaining fixed. The scale-normalised intensity
\(\hat\lambda_{i,t}/\hat\sigma_{i,t}\) is predominantly negative across
stocks and dates. This is consistent with the empirically relevant
orientation in which the nonlinear feedback increases central
concentration and contributes to the peak--shoulder--tail deformation
relative to the Gaussian benchmark. The smoothed paths also show
stock-specific variation in the persistence and magnitude of this
distortion, supporting the interpretation of the MBMM parameters as
daily tail-shape state variables rather than as static stock
characteristics.

To interpret what drives the high realised kurtosis on this day, we
decompose the fitted MBMM density into peak, shoulder, and tail
components relative to the Gaussian benchmark. The area measures are
diagnostic excess-mass components: \(A_{\mathrm{peak}}\) measures
central excess density, \(A_{\mathrm{shoulder}}\) measures depletion in
the intermediate-return region, and \(A_{\mathrm{tail}}\) measures
outer-tail excess.

Table~\ref{tab:same_day_area_components} shows that, for all ten
stocks, the largest component is \(A_{\mathrm{shoulder}}\). 
\begin{table}[!htbp]
\centering
\caption{Same-day MBMM peak--shoulder--tail decomposition on 9 April 2025}
\label{tab:same_day_area_components}
\scriptsize
\setlength{\tabcolsep}{4.2pt}
\renewcommand{\arraystretch}{1.12}
\begin{tabular*}{\textwidth}{@{\extracolsep{\fill}}lrrrrrrrrr@{}}
\toprule
Stock
& \(RK\)
& \(c\)
& \(\alpha\)
& \(z_P\)
& \(z_T\)
& \(A_{\mathrm{peak}}\)
& \(A_{\mathrm{shoulder}}\)
& \(A_{\mathrm{tail}}\)
& \(\Delta^{\mathrm{obj}}\) \\
&
&
&
&
&
&
&
&
&
\((\%)\) \\
\midrule
AAPL  & 26.110 & 0.178 & -0.462 & 1.179 & 2.792 & 0.177 & 0.277 & 0.000 & 89.7 \\
AMZN  & 31.982 & 0.262 & -0.566 & 1.015 & 2.302 & 0.209 & 0.319 & 0.003 & 85.9 \\
GOOGL & 40.893 & 0.208 & -0.624 & 1.109 & 2.579 & 0.241 & 0.350 & 0.001 & 76.0 \\
JPM   & 30.628 & 0.241 & -0.477 & 1.048 & 2.398 & 0.176 & 0.279 & 0.002 & 73.1 \\
META  & 20.435 & 0.321 & -0.495 & 0.937 & 2.079 & 0.174 & 0.277 & 0.005 & 83.3 \\
MSFT  & 34.195 & 0.315 & -0.746 & 0.944 & 2.098 & 0.272 & 0.386 & 0.006 & 75.3 \\
NVDA  & 20.114 & 0.228 & -0.464 & 1.071 & 2.466 & 0.172 & 0.274 & 0.001 & 84.6 \\
TSLA  & 21.066 & 0.100 & -0.295 & 1.462 & 3.731 & 0.116 & 0.180 & 0.000 & 76.3 \\
UNH   & 12.910 & 0.343 & -0.521 & 0.912 & 2.011 & 0.181 & 0.286 & 0.006 & 82.6 \\
XOM   & 34.204 & 0.263 & -0.506 & 1.013 & 2.295 & 0.184 & 0.290 & 0.003 & 76.3 \\
\bottomrule
\end{tabular*}

\vspace{0.15cm}

\begin{minipage}{0.97\textwidth}
\footnotesize
\textit{Notes:}
The table reports the fitted MBMM parameters and diagnostic
peak--shoulder--tail area components for the same stock-day shown in
Figure~\ref{fig:same_day_distribution_fits}. The area measures are
computed relative to the Gaussian benchmark. \(\alpha=\lambda/\sigma\),
and \(\Delta^{\mathrm{obj}}\) is the percentage reduction in the
quantile-distance objective relative to GBM.
\end{minipage}
\end{table}
Thus, the
high-kurtosis shape observed on 9 April 2025 is not explained only by
outer tails. It is mainly a redistribution of probability mass away
from intermediate returns, with corresponding excess mass concentrated
near the centre and, to a smaller extent, in the tails. The tail component is small for most stocks because the fitted
tail-transition point \(z_T\) remains relatively far from the centre,
especially for TSLA and AAPL. By contrast, stocks with a lower
\(z_T\) and stronger distortion intensity, such as MSFT, UNH, and
META, show a more visible outer-tail component. 

The main empirical
message is therefore simple: on this high-kurtosis day, MBMM improves
the fit because it captures the full peak--shoulder--tail geometry,
not merely because it adds fatter tails to the Gaussian benchmark.

\subsection{Are the MBMM shape parameters forecastable?}
\label{sec:forecasting_tail_indicators}

The previous analysis treats the estimated MBMM parameters as daily
descriptors of intraday return-shape geometry. We now ask a distinct
question: do these estimated parameter sequences contain exploitable
one-step-ahead temporal structure? To answer this question, we assess if a simple first-order dynamic specification systematically outperforms
a no-change forecast out of sample \citep[e.g. see][]{darrat2000testing,moosa2016random}.

We consider three forecasting targets. The two primary targets are the
log-transformed shape parameter,
\(
x^{(c)}_{i,t}=\log(\widehat c_{i,t}),
\)
and the raw distortion parameter,
\(
x^{(\lambda)}_{i,t}=\widehat\lambda_{i,t}.
\)
Because $\widehat\lambda_{i,t}
=\widehat\alpha_{i,t}\widehat\sigma_{i,t}$ combines distortion intensity
and scale, we additionally consider the volatility-normalised companion
target
\(
x^{(\alpha)}_{i,t}
=
\widehat\alpha_{i,t}
=
\frac{\widehat\lambda_{i,t}}{\widehat\sigma_{i,t}}.
\) 
Forecasts are generated separately for each stock using a expanding-window design. Let $T_i$ denote the number of available
observations for stock $i$. The first $\lfloor T_i/2\rfloor$ observations
form the initial estimation sample. At each forecast origin $t$, the model
is estimated using all observations available through $t$, a one-step-ahead
forecast for $t+1$ is produced, and the estimation window is then expanded
by one trading day. 

The benchmark is a driftless random walk (RW), under which the next
parameter estimate is forecast by its current value,
\begin{equation}
\widehat{x}^{(m),\mathrm{RW}}_{i,t+1|t}
=
x^{(m)}_{i,t},
\qquad
m\in\{c,\lambda,\alpha\}.
\label{eq:tail_forecast_rw}
\end{equation}
We compare the RW forecast with a stock-specific AR(1),
\begin{equation}
x^{(m)}_{i,t+1}
=
a_i^{(m)}
+
\phi_i^{(m)}x^{(m)}_{i,t}
+
\eta^{(m)}_{i,t+1},
\qquad
m\in\{c,\lambda,\alpha\},
\label{eq:tail_forecast_ar1}
\end{equation}
estimated recursively using the same expanding-window scheme. The
single-lag specification is intentional, since the objective is not to search
over alternative time-series models, but to determine whether even a
minimal dynamic model can improve on the no-change forecast. 

Forecast performance is evaluated stock by stock and then aggregated
giving equal weight to each stock. Let $\mathcal O_i$ denote the set of
out-of-sample forecast dates for stock $i$, with
$T_i^{\mathrm{OOS}}=|\mathcal O_i|$. For forecasting model $a$, the
stock-specific RMSE is
\begin{equation*}
\mathrm{RMSE}^{(m)}_i(a)
=
\left[
\frac{1}{N_i^{\mathrm{OOS}}}
\sum_{t+1\in\mathcal O_i}
\left(
x^{(m)}_{i,t+1}
-
\widehat{x}^{(m),a}_{i,t+1|t}
\right)^2
\right]^{1/2},
\label{eq:tail_forecast_rmse_stock}
\end{equation*}
and the reported RMSE is the average across the $N$ stocks. The MAE loss is constructed analogously. Equal weighting prevents stocks with
longer out-of-sample histories from mechanically receiving greater
importance in the reported accuracy measures. As an additional accuracy measure, we also report directional accuracy \citep{blaskowitz2011economic,costantini2016forecasting,costantini2026use}. Let
\[
D^{(m)}_{i,t+1}
=
\operatorname{sgn}
\left(
x^{(m)}_{i,t+1}-x^{(m)}_{i,t}
\right)
\]
denote the realised direction of change and
\[
\widehat D^{(m)}_{i,t+1|t}
=
\operatorname{sgn}
\left(
\widehat x^{(m),\mathrm{AR(1)}}_{i,t+1|t}
-
x^{(m)}_{i,t}
\right)
\]
the corresponding predicted direction. Stock-specific directional
accuracy is
\begin{equation*}
\mathrm{DA}^{(m)}_i
=
100
\times
\frac{1}{N_i^{\mathrm{OOS}}}
\sum_{t+1\in\mathcal O_i}
\mathbf 1
\left[
\widehat D^{(m)}_{i,t+1|t}
=
D^{(m)}_{i,t+1}
\right],
\label{eq:tail_directional_accuracy_stock}
\end{equation*}
and the reported DA is the equally weighted average across stocks. Directional accuracy is not reported for the RW because it mechanically
predicts zero change and therefore contains no directional signal. Finally, to test whether the reduction in squared forecast error is
statistically detectable, we report thew results of the HAC-robust \cite{DieboldMariano1995} test.

\color{black}The Panel A of Table~\ref{tab:for_res} reports the
out-of-sample forecasting results. 
\begin{table}[!htbp]
\centering
\small
\setlength{\tabcolsep}{6pt}
\renewcommand{\arraystretch}{1.15}
\caption{Out-of-sample forecastability of MBMM shape and distortion parameters}
\label{tab:for_res}
\begin{threeparttable}

\begin{tabular}{llrrrr}
\toprule

\multicolumn{6}{l}{\textit{Panel A: Out-of-sample forecast accuracy}} \\
\addlinespace[2pt]
Target & Model & RMSE & MAE & Gain vs RW (\%) & DA (\%) \\
\midrule

$\log(\widehat c)$
& RW & 0.5404 & 0.3172 & -- & -- \\
& AR(1) & 0.3863 & 0.2130 & 28.51 & 74.32 \\

\addlinespace[2pt]

$\widehat\lambda$
& RW & 0.000356 & 0.000197 & -- & -- \\
& AR(1) & 0.000268 & 0.000152 & 24.78 & 70.16 \\

\addlinespace[2pt]

$\widehat\alpha=\widehat\lambda/\widehat\sigma$
& RW & 0.2872 & 0.1924 & -- & -- \\
& AR(1) & 0.2189 & 0.1460 & 23.77 & 71.28 \\

\addlinespace[6pt]
\midrule

\multicolumn{6}{l}{
\textit{Panel B: Diebold--Mariano HAC squared-loss test, RW versus AR(1)}
} \\
\addlinespace[2pt]
Target
& \multicolumn{2}{c}{Mean loss differential}
& HAC $t$-stat.
& \multicolumn{2}{c}{One-sided $p$} \\
\midrule

$\log(\widehat c)$
& \multicolumn{2}{c}{0.147756}
& 7.404
& \multicolumn{2}{c}{$<0.001$} \\

$\widehat\lambda$
& \multicolumn{2}{c}{5.609e-08}
& 3.132
& \multicolumn{2}{c}{$<0.001$} \\

$\widehat\alpha=\widehat\lambda/\widehat\sigma$
& \multicolumn{2}{c}{0.034179}
& 9.900
& \multicolumn{2}{c}{$<0.001$} \\

\bottomrule
\end{tabular}
\end{threeparttable}
\renewcommand{\arraystretch}{1.00}
\end{table}
The AR(1) improves one-step-ahead forecast accuracy substantially for all
three parameter sequences. For $\log(\widehat c)$, RMSE decreases from
0.5404 under the RW to 0.3863 under the AR(1), corresponding to a
28.51\% reduction. The improvement is 24.78\% for raw
$\widehat\lambda$ and 23.77\% for the volatility-normalised
$\widehat\alpha$. The same ranking is obtained using MAE, indicating that
the reductions in RMSE are not driven solely by a small number of large
forecast errors.
In terms of DA, the AR(1) also predicts the direction of parameter changes very well, as DA equals 74.32\% for
$\log(\widehat c)$, 70.16\% for $\widehat\lambda$, and 71.28\% for
$\widehat\alpha$.  Statistical evidence on forecast
improvement is reported in
Panel B of Table \ref{tab:for_res}, and suggest that differences in Panel A are statistically significant at 99\% confidence level. Therefore, we can conclude that the shape-geometry parameters are forecastable quantities.

\color{black}

\section{Discussion and conclusions}
\label{sec:conclusions}

This paper studies whether the shape of intraday return distributions can be
measured in a way that complements realised kurtosis. Realised kurtosis provides
an informative measure of the overall magnitude of fourth-moment risk, but it
compresses the geometry of an entire return distribution into a single scalar.
Consequently, trading days characterised by similar realised kurtosis need not
exhibit the same allocation of probability mass across the \textcolor{black}{peak}, intermediate
returns, and the tails. The purpose of this paper is to provide a framework for
measuring and interpreting these differences.

Our analysis builds on the modified Brownian motion model (MBMM) proposed by
\cite{dhesi2016modelling,DhesiShakeelXiao2016}. The MBMM provides a particularly
useful technical foundation for this purpose because it represents returns as a
nonlinear deformation of a Gaussian innovation. The associated feedback
function separates the location of the deformation from its intensity and
therefore provides a natural mechanism through which different return-shape
configurations can be generated. Starting from this representation, we reinterpret
the MBMM parameters and the underlying nonlinear feedback in explicitly
geometrical terms.

The methodological contribution of the paper is therefore threefold. First, we
show how the MBMM feedback mechanism can be used to characterise the geometry
of a return distribution rather than being interpreted only as a device for
generating non-Gaussian returns. In particular, we introduce a
peak--shoulder--tail decomposition that measures where the fitted MBMM density
gains or loses probability mass relative to the Gaussian benchmark. The
resulting central-excess, shoulder-depletion, and outer-tail-excess components
describe how probability mass is redistributed over the return distribution.
We complement this decomposition with two feedback landmarks, the
peak-distortion point \(z_P(c)\) and the tail-transition point \(z_T(c)\), which
provide an interpretation of how the shape parameter \(c\) determines the
location of the nonlinear deformation on the latent innovation scale. These
landmarks are distinct from the density crossing points defining the
peak--shoulder--tail regions: the former describe the mechanism generating the
deformation, whereas the latter describe its consequences for the fitted return
density.

Second, we propose a simulation-based quantile minimum-distance estimator for
the MBMM parameters. Previous implementations of the model relied on
simulation-based goodness-of-fit procedures applied to binned return
distributions. Our estimator instead compares empirical and model-implied
quantiles over a fixed probability grid and therefore avoids dependence on
histogram construction while exploiting information from the centre, shoulders,
and tails of the distribution simultaneously. This makes the estimation
procedure particularly suitable for repeated estimation on relatively short
within-day samples. The Monte Carlo analysis investigates the finite-sample
behaviour of this estimator and provides evidence on its ability to recover the
underlying MBMM parameters under nonlinear data-generating processes while
suppressing the nonlinear component when the Gaussian benchmark is the true
data-generating process.

Third, we move the MBMM framework to a stock-day measurement setting by using
high-frequency intraday observations. Rather than estimating a single return
distribution over a long observation period, we estimate the model separately
for each trading day using one-minute returns. This produces daily estimates of
the MBMM geometry and associates each observed value of realised kurtosis with
a corresponding fitted return-shape configuration. The resulting objects can
therefore be interpreted as daily state variables describing the geometry of
intraday return risk.

\color{black}

From the perspective of investors and risk managers, the distinction between
realised kurtosis and MBMM geometry developed in this paper is operationally
relevant. Two stock-days can exhibit very similar realised kurtosis, that is,
\(RK_t \simeq RK_s\) for two days \(t\) and \(s\), while implying materially
different allocations of probability mass across the \textcolor{black}{peak}, shoulders, and
tails. One day may be characterised by pronounced central concentration and
shoulder depletion, whereas another may exhibit a transition closer to the
centre and a relatively larger outer-tail excess. Hence, the same fourth-moment
magnitude need not correspond to the same distributional risk configuration.
This distinction may be useful for threshold-based risk monitoring, the
diagnostic interpretation of Value-at-Risk and Expected Shortfall,
stress-scenario design, intraday exposure limits, and the assessment of whether
changes in tail risk arise primarily from intermediate-return depletion or from
a stronger outer-tail component. In this sense, the proposed geometry
complements conventional scalar risk measures by describing how tail risk is
formed rather than only how large it is. We stress, however, that the proposed
geometry is not a trading signal and the present analysis does not establish
profitability or portfolio gains. It should instead be interpreted as a
distributional risk-shape diagnostic that complements realised volatility,
realised kurtosis, and conventional tail-risk measures.

\color{black}

We evaluate the proposed framework through both simulation and an empirical
application. The empirical analysis uses regular-session one-minute returns for
ten liquid U.S. stocks from May 2021 to April 2026. Two main empirical findings
emerge.

First, the fitted MBMM provides a substantially better in-sample representation
of the intraday return distribution than the Gaussian geometric Brownian motion
benchmark. The reduction in the quantile-distance objective is large across all
stocks and is positive for every estimator-ready stock-day. This result is
important for the interpretation developed in the paper: the proposed geometry
is not constructed from an arbitrary deformation that bears little relation to
the observed data. Rather, the estimated MBMM provides an empirically relevant
representation of the intraday return distribution and captures features of its
central concentration, intermediate-return depletion, and tail behaviour that
are systematically missed by the Gaussian benchmark. The resulting
peak--shoulder--tail decomposition can therefore be interpreted as a
data-supported description of the fitted intraday return shape. This remains an
in-sample distributional-fit result and should not be interpreted as evidence of
forecasting or trading performance.

Second, the MBMM-derived quantities display substantial temporal persistence and
behave as forecastable daily state variables. In the out-of-sample
indicator-forecasting exercise, both the expanding-mean and AR(1) specifications
substantially outperform the random-walk benchmark. The improvement of the
AR(1) over the expanding-mean benchmark is considerably smaller, particularly
for the shape component. The evidence therefore suggests that the estimated
geometry is not merely day-specific estimation noise, but neither does it
display particularly rich short-run autoregressive dynamics. A more appropriate
interpretation is that the MBMM-derived quantities behave as persistent state
variables evolving around slowly changing levels.

This distinction is also important for the scope of the contribution. The
present paper develops a measurement and estimation framework for the geometry
of intraday return distributions and establishes that the resulting indicators
contain persistent and forecastable structure. We do not establish that
forecasting these quantities generates profitable trading strategies. Their
predictability nevertheless motivates a natural next stage of research:
developing dedicated dynamic models for the MBMM geometry and investigating
whether forecasts of these states contain incremental information for future
returns, volatility, tail events, portfolio decisions, or other economically
relevant outcomes. Such an analysis would require a separate out-of-sample
economic evaluation and should therefore be regarded as an extension rather
than as a conclusion of the present study.

The analysis is subject to several limitations. The empirical application is
based on ten highly liquid U.S. equities and regular-session one-minute returns,
and the interpretation of the daily MBMM states necessarily depends on the
quality and stability of this intraday measurement layer. Moreover, although
the proposed estimator is evaluated through Monte Carlo experiments, more
extensive investigation of its behaviour under alternative data-generating
processes, sampling frequencies, and weighting schemes would be useful. Future
research could therefore extend the analysis to a larger equity universe,
alternative asset classes, different intraday sampling frequencies, and
alternative quantile-weighting schemes. Another important direction is to
investigate whether the proposed geometry improves the measurement and
forecasting of portfolio risk, Value-at-Risk, Expected Shortfall, or
state-dependent stress scenarios. More generally, modelling the joint dynamics
of the peak, shoulder, and tail components may provide a way to study not only
whether tail risk changes over time, but also how the internal geometry through
which that risk is generated evolves.

\section*{Acknowledgment}

This work was partially supported by the project “Societal and Economic Resilience within multi-hazards environment in Romania” funded by European Union – NextgenerationEU and Romanian Government, under National Recovery and Resilience Plan for Romania, contract no.760050/ 23.05.2023, cod PNRR-C9-I8-CF 267/ 29.11.2022, through the Romanian Ministry of Research, Innovation and Digitalization, within Component 9, Investment I8. This work was also supported by the project “A better understanding of socio-economic systems using quantitative methods from Physics” funded by the European Union – NextgenerationEU and the Romanian Government, under National Recovery and Resilience Plan for Romania, contract no 760034/23.05.2023, cod PNRR-C9-I8-CF255/29.11.2022, through the Romanian Ministry of Research, Innovation and Digitalization, within Component 9, Investment I8.

\bibliographystyle{apalike}
\bibliography{rQUFguide}

\end{document}